\documentclass[preprint,prd,superscriptaddress,nofootinbib]{revtex4-1}

\usepackage{natbib}

\usepackage[utf8]{inputenc}
\usepackage{dcolumn} 
\usepackage{graphicx}
\usepackage{slashed,xcolor}
\usepackage{amsmath,amsfonts,bm}
\usepackage{amssymb}
\usepackage{soul}
\newcommand{\beq}{\begin{equation}}
\newcommand{\eeq}{\end{equation}}
\newcommand{\bea}{\begin{eqnarray}}
\newcommand{\eea}{\end{eqnarray}}
\newcommand{\qedl}{${\rm QED}_L$ }
\newcommand{\QEDL}{{\rm QED}_L}

\def\embr{\color{brown}}

\newcommand{\VTC}{V_{\rm TC}}
\newcommand{\VCTC}{V_{\overline{\raisebox{2.3 mm}{}\mathrm{TC}}}}
\newcommand{\KCTC}{K_{\overline{\raisebox{2.3 mm}{}\mathrm{TC}}}}

\newcommand{\MCTC}{M_{\overline{\raisebox{2.3 mm}{}\mathrm{TC}}}}
\newcommand{\dCTC}{\delta^{\overline{\raisebox{2.3 mm}{}\mathrm{TC}}}}
\newcommand{\dCTCZ}{\delta^{\overline{\raisebox{2.3 mm}{}\mathrm{TC}}0}}

\newcommand{\VQEDL}{V_{\mathrm{QED}_L}}

\newcommand{\ba}{\begin{eqnarray}}
\newcommand{\ea}{\end{eqnarray}}

\newcommand{\be}{\begin{equation}}
\newcommand{\ee}{\end{equation}}

\begin{document}

\newcommand*{\CU}{Physics Department, Columbia University, New York City, New York 10027, USA}\affiliation{\CU}
\newcommand{\innovation}{Collaborative Innovation Center of Quantum Matter, Beijing 100871, China}
\newcommand{\chep}{Center for High Energy Physics, Peking University, Beijing 100871, China}
\newcommand{\pkuphy}{School of Physics, Peking University, Beijing 100871,
China}
\newcommand{\KeyLab}{State Key Laboratory of Nuclear Physics and Technology,
Peking University, Beijing 100871, China}

\title{$\pi-\pi$ scattering, QED and finite-volume quantization}
\author{Norman Christ}\affiliation{\CU}
\author{Xu Feng}\affiliation{\pkuphy}\affiliation{\innovation}\affiliation{\chep}\affiliation{\KeyLab}
\author{Joseph Karpie}\affiliation{\CU}
\author{Tuan Nguyen}\affiliation{\CU}

\date{October 17, 2021}

\begin{abstract}  Using the Coulomb gauge formulation of QED we present a lattice QCD procedure to calculate the $\pi^+\pi^+$ scattering phase shift including the effects of the Coulomb potential which appears in this formulation.  The approach described here incorporates the effects of relativity and avoids finite-volume corrections that vanish as a power of the volume in which the lattice calculation is performed.  This is the first step in developing a complete lattice QCD calculation of the electromagnetic and isospin-breaking light-quark mass contributions to $\varepsilon'$, the parameter describing direct CP violating effects in $K_L\to\pi\pi$ decay. 
\end{abstract}

\maketitle

\tableofcontents

\newpage

\section{Introduction}

The $K\to\pi\pi$ decays are important to our understanding of CP violation from the Weak interaction. Indirect CP violation in the $K_L\to\pi\pi$ decay arises from the admixture of the CP-even combination of $K^0$ and $\overline{K^0}$ mesons in the long-lived $K_L$ state. It is described by the parameter $\varepsilon$ whose measured magnitude is $2.228(11)\times10^{-3}$. Direct CP violation arises from the CP-odd component of $K_L$, which can also directly decay into two pions. It is described by the parameter $\varepsilon'$ whose value is 3 orders of magnitude smaller than $\varepsilon$.  Due to its small size, direct CP violation in $K\to\pi\pi$ decay, represented by the parameter $\varepsilon'$, is very sensitive to new physics.

The experimental measurement determines the direct CP-violating ratio $\operatorname{Re}(\varepsilon'/\varepsilon)=16.6(2.3)\times10^{-4}$~\cite{Batley:2002gn, PhysRevD.83.092001}. The current standard model prediction for this quantity computed using lattice QCD is given by $\operatorname{Re}(\varepsilon'/\varepsilon) = 21.7(2.6)(6.2)(5.0)\times10^{-4}$~\cite{Abbott:2020hxn}.   Here the first error is statistical, the second is systematic and the third arises from the neglect of electromagnetism and the effects of the isospin-violating mass difference, $m_u-m_d$ (which will be referred to below as $m_u-m_d$ effects) that are the topic of this paper. 

For most processes, electromagnetic corrections are on the order of the fine structure constant $\alpha=1/137$.  However, the quantity $\varepsilon'/\varepsilon$ involves with equal weight the amplitudes $A_0$ and $A_2$ for the decay of a $K^0$ meson into two pions in the isospin 0 and isospin 2 $\pi\pi$ states, respectively. Since $A_2$ is suppressed relative to $A_0$ by a factor of 22, a feature called the $\Delta I=1/2$ rule, the electromagnetic modifications to $A_0$ can in principle induce corrections to $A_2$ which are 22 times larger than this usual $1/137$ scale.   Such effects can then propagate into the standard model prediction for $\operatorname{Re}(\varepsilon'/\varepsilon)$.  

These electromagnetic effects have been extensively studied using chiral perturbation theory~\cite{Cirigliano:1999ie, Cirigliano:1999hj, Cirigliano:2000zw, Wolfe:2000rf, Cirigliano:2003gt, Cirigliano:2019cpi}.  In fact, it was the recent estimate of these isospin-breaking effects given in Ref.~\cite{Cirigliano:2019cpi} that was used in Ref.~\cite{Abbott:2020hxn} and appears above as an estimate of the error resulting from the neglect of these effects.   Given the difficulty of this calculation and the large size of these corrections to $\operatorname{Re}(\varepsilon'/\varepsilon)$, an ab initio calculation of these effects using lattice QCD would be of value.  In this paper we outline a possible strategy for including electromagnetic and $m_u-m_d$ effects in a lattice QCD calculation of $\varepsilon'$ and work out in detail the first step in this strategy.  The approach presented here builds upon that proposed in Ref.~\cite{Christ:2017pze}.  See also the related paper of Cai and Daviodi~\cite{Cai:2018why}.

There are a number of important challenges that must be addressed if electromagnetic and $m_u-m_d$ effects  are to be computed using lattice methods:
\begin{enumerate}
\item The calculation of $K\to\pi\pi$ depends heavily on the finite-volume methods of L\"uscher~\cite{Luscher:1990ux} and Lellouch and L\"uscher~\cite{Lellouch:2000pv} which rely on the exponentially-localized finite-range interactions of QCD.   Adding electricity and magnetism (E\&M) introduces long-range interactions, inconsistent with the Lellouch-L\"uscher strategy.  This problem is dramatically illustrated by the fact that the scattering phase shifts which play a central role in the finite-volume treatment of Lellouch and L\"uscher are not even defined when E\&M effects are included, with the long-distance wave functions acquiring phases which grow logarithmically with distance, $\sim\eta\ln(kr)$ where $k$ is the center-of-mass (CoM) momentum of the scattering particles, $r$ the CoM separation of the outgoing particles and $\eta = me^2/(4\pi k)$ is the Sommerfeld parameter with $\pm e$ their charge and $m$ their mass.
\item The usual treatment of $K\to\pi\pi$ decay relies on isospin symmetry to distinguish two independent $\pi\pi$ final states, one with $I=0$ and the other with $I=2$.  Electromagnetism and $m_u-m_d$ effects break isospin symmetry, allowing the $\pi\pi$ states with $I=0$ and $I=2$ to mix.  As a result, the final state scattering that is part of the $K\to\pi\pi$ decay becomes a coupled, two-channel problem, requiring a more complex treatment of both the finite-volume eigenstates and the infinite-volume decay processes.
\item A process such as $K\to\pi\pi$ decay, which involves the acceleration of charge, will contain well-known infrared singularities~\cite{Bloch:1937, Yennie:1961ad, Lee:1964is} which are removed by a careful treatment of the possible, near-degenerate final states, which include the intended $\pi\pi$ state as well as states with one or more emitted photons in addition to the two pions.  While the effects of such soft radiation can be computed using standard methods for the case of the infinite-volume decay, possible photon emission in a finite-volume lattice calculation may introduce additional complications making the already challenging two-channel problem described above into a problem with four channels, two of which are three-particle channels.
\end{enumerate}

We intend to carry out this calculation of the first-order E\&M and $m_u-m_d$ contributions to $\varepsilon'$ by exploiting the linear character of such a first-order calculation.  We will separate the problem into simpler pieces which can be computed independently and then added together to obtain the final result.  The first step in this divide-and-conquer approach treats the effects of electromagnetism in the Coulomb or radiation gauge.  As is reviewed in the next section, in the Coulomb gauge, the E\&M vector potential $\vec A$ is required to be transverse, $\vec \nabla \cdot\vec A=0$, and the resulting E\&M Hamiltonian divides into two separate terms.  The first is the familiar $1/|\vec r\,|$ Coulomb potential between two charge density operators evaluated at equal times and separated by the displacement $\vec r$.  The second, independent piece involves the coupling of transverse photons to the spatial components of the E\&M current operator.  These two components describe different physical phenomena in the rest frame of the kaon and their effects can be treated separately.  Since, in a lattice calculation, we are fixed into a frame and will naturally work in a finite volume in which the kaon is at rest, no added difficulties are introduced by the choice of such a  Lorentz non-covariant gauge. 

The Coulomb potential gives the largest effect of electromagnetism when the charged pions are moving at non-relativistic velocities; effects which are familiar from non-relativistic quantum mechanics.  Its long-range, $1/r$ character complicates the usual finite-volume methods of lattice QCD and introduces singular effects resulting from the scattering of charged particles at long distances, including the logarithms of $r$ present in the asymptotic behavior of scattering solutions.  It is this Coulomb potential component of the E\&M problem that we will discuss in this paper.

The second component of the Coulomb-gauge Hamiltonian is the photon-current interaction which emits and absorbs massless photons.  This component is the source of the usual infrared divergences but, as with the Coulomb interaction, also includes singular short-distance effects which require careful treatment.  This part of the problem is not treated in the present paper.  However, for both the Coulomb and radiation parts, the long-distance effects which may be incompatible with the finite volumes used in lattice QCD are fundamentally classical.  This suggests that these long-distance effects can be separated and computed analytically, leaving the portion of the problem to be treated using lattice methods free of these long distance difficulties.

For the case of the Coulomb potential, this separation of short- and long-distance effects is easily achieved by writing the Coulomb potential as the sum of two pieces: 
\begin{equation}
\frac{e^2}{4\pi|\vec r|} = \frac{e^2}{4\pi|\vec r|}\theta(R-r) +  \frac{e^2}{4\pi|\vec r|}\theta(r-R)
\label{eq:truncate}
\end{equation}
where we refer to the fixed separation $R$ as the truncation radius and the two terms on the right-hand side of Eq.~\eqref{eq:truncate} as the short-distance truncated Coulomb potential $\VTC(r)$ (the left-hand term) and the long-distance compliment of the truncated Coulomb potential $\VCTC(r)$ (the right-hand term).  Given the linearity of the first order correction that we wish to compute, we can treat these two terms separately.  The effects of $\VTC(r)$ can be directly combined with those of QCD in a lattice calculation provided we choose $2R$ to be smaller than the linear size of the volume used in this QCD+QED calculation.  As we will show, the second term can be treated analytically if $R$ is sufficiently large that terms which decrease exponentially for large $R$ can be neglected.

In the present paper, we treat a portion of the Coulomb potential problem, studying $\pi^+\pi^+$ scattering and derive results which can be combined with a lattice QCD calculation to obtain the $\pi^+\pi^+$ scattering phase shift accurate to first order in $\alpha$, including both the effects of QCD and the full Coulomb potential.  As we will show, this combined lattice QCD and analytic calculation can be carried out with the only finite-volume errors being those which vanish exponentially in the linear extent of the volume used for the lattice QCD calculation.  In a practical use of our result, there will be power-law corrections that result from the usual neglect of phase shifts for those partial waves with angular momentum $l$ larger than some minimum value.  The remaining part of the Coulomb problem needed for the calculation of $\varepsilon'$ requires examining the two-channel $\pi^+\pi^-$ and $\pi^0\pi^0$ scattering and the effects of the Coulomb potential on the actual $K\to\pi\pi$ decay.  This problem has been treated in the non-relativistic case in our earlier proceedings~\cite{Christ:2017pze} and we plan to provide a solution to this problem including relativistic effects in a later paper.

The second part of the problem of computing the E\&M and $m_u-m_d$ contributions to $\varepsilon'$ requires determining the effects of the transverse radiation.  In a fashion similar to our approach to the Coulomb potential, we plan to divide the transverse photons into two groups whose energies lie above or below a boundary energy $E_B$.  Those photons with energy $E_\gamma < E_B$ are treated classically using the usual Bloch-Nordsiek methods while those whose energy $E_\gamma > E_B$ will be treated explicitly using lattice QCD in a finite volume with linear extent $L$.  The correctness of the classical treatment requires that $E_B/\Lambda_\mathrm{QCD} \ll 1$ so that structure-dependent effects can be neglected.  At the same time, the accurate treatment of the hard radiation using finite-volume lattice QCD requires $1/L \ll E_B$.  These combined requirements will result in the neglect of corrections behaving as a power of $1/(\Lambda_\mathrm{QCD}L)$. This treatment of the radiation part of the E\&M problem is our long term strategy and the focus of current study.

Our use of the truncated Coulomb potential which allows us to include QED in a finite-volume lattice calculation is different from the more conventional approach of called \qedl~\cite{Hayakawa:2008an}.  If specialized to the Coulomb potential alone, the \qedl approach would express the $1/r$ Coulomb potential in a finite spatial volume of size $L^3$ as a conventional periodic Fourier series over wave numbers $\vec k = 2\pi(n_1,n_2,n_3)/L$ for integers $\{n_i\}_{1 \le i \le 3}$ from which the ill-defined term with $\vec k = (0,0,0)$ has been omitted.  When beginning this project, we examined this approach to QED in a finite volume and present some of our results in appendices to this paper.   

Here our results are closely related to those of Beane and Savage~\cite{Beane:2014qha} and Beane {\it et al.}~\cite{NPLQCD:2020ozd}.  However, the treatment presented in Appendix~\ref{sec:QEDL-quantization} may provide a useful compliment to this earlier work since it does not involve an effective range approximation to the energy dependence of the scattering phase shift.  We choose to use the truncated Coulomb potential because this position-space approach appears easier to understand and the absence of new power-law finite-volume corrections may be an important advantage in a calculation in which power-law finite-volume effects are being exploited to determine the scattering phase shifts.  In Appendix~\ref{sec:QEDL-numerics}, we investigate the size of the power-law finite-volume corrections to the scattering phase shifts determined numerically in the case of non-relativistic quantum mechanics with a simple scattering potential and find a large $1/L$ correction which gives an easy-to-correct energy shift and a small upper bound on the $1/L^3$ corrections, suggesting that the \qedl approach may also work well for the problem at hand.

The paper is organized as follows. In Sec.~\ref{sec:formulation}, we recall the Coulomb gauge treatment of QED  and present the proposed method to compute the scattering of two identical, charged spin-zero particles including the combined effects of QCD and the instantaneous Coulomb potential where the latter is included only to first order in $\alpha$.  The usual finite-volume quantization method that can be used to determine the $O(\alpha)$ contributions of $\VTC$ to the $\pi^+\pi^+$ scattering phase shift are described in Sec.~\ref{sec:VTC} while in Sec.~\ref{sec:VCTC} we present the details of an analytic calculation of the $O(\alpha)$ contributions of $\VCTC$ to this phase shift.  The analytic results of this section are expressed in terms of on-shell properties of QCD specified by the $\pi^+\pi^+$ scattering phase shift in the absence of electromagnetic corrections and the $\pi^+$ electromagnetic form factor.   Finally, conclusions and future plans are described in Sec.~\ref{sec:conclusion}.  The paper has two appendices, \ref{sec:QEDL-quantization} and \ref{sec:QEDL-numerics}, whose contents are described above.

\section{Formulation and strategy}
\label{sec:formulation}

As described above, we propose to compute the QED corrections to $\pi^+\pi^+$ scattering by using the Coulomb-gauge formulation of QED.  In that approach the Minkowski-space QED Lagrangian is written as
\begin{eqnarray}
L_{\mathrm{EM}} &=& \frac{1}{2} \int d^3 r \left\{ \left(\partial_t \vec A(\vec r)\right)^2 - \left(\vec \nabla \times \vec A(\vec r)\right)^2 + \vec j(\vec r)\cdot\vec A(\vec r) \right\} \label{eq:coulomb} \\
&&\hskip 1.5 in  -\frac{1}{2} \int d^3 r d^3 r^\prime \rho(\vec r)\frac{1}{4\pi|\vec r - \vec r\,^\prime|}\rho(\vec r\,^\prime), \nonumber
\end{eqnarray}
a standard textbook result~\cite{Srednicki:2007qs} for the quantum treatment of the electromagnetic field.  Here $\vec j(\vec r)$ and $\rho(\vec r)$ are the current and charge density operators for the quarks to which the E\&M field couples.   This is the treatment of QED that is used in the lattice calculation described in Sec.~\ref{sec:VTC}.  However, in Sec.~\ref{sec:VCTC} where we study the E\&M interations of pions at large distances, we will use ``scalar'' QED and $\vec j(\vec r)$ and $\rho(\vec r)$ will be written in terms of the pion field and its spatial and temporal derivatives.  Since we plan to compute E\&M corrections to first order in $\alpha$, the two interaction terms on the right-hand side of Eq.~\eqref{eq:coulomb} can be treated independently and the results simply added together in the end.  This will allow us to consider separately the Coulomb interaction with its long-range distortion of the two-particle scattering problem and the interaction $\vec j\cdot\vec A$ of the transverse photons which requires the Bloch-Nordsiek treatment~\cite{Bloch:1937}.  In this paper we will focus on the corrections arising from the second term, the Coulomb interaction.

In our next step this Coulomb potential is itself further divided into two terms $\VTC$ and $\VCTC$:
\begin{eqnarray}
\VTC &=&  \frac{1}{2} \int d^3 r d^3 r^\prime \rho(\vec r)\,\frac{\theta\bigl(R-|\vec r-\vec r\,'|\bigr)}{4\pi|\vec r - \vec r\,^\prime|}\rho(\vec r\,^\prime) 
\label{eq:VTCop} \\
\VCTC &=&  \frac{1}{2} \int d^3 r d^3 r^\prime \rho(\vec r)\,\frac{\theta\bigl(|\vec r-\vec r\,'|-R\bigr)}{4\pi|\vec r - \vec r\,^\prime|}\rho(\vec r\,^\prime) \label{eq:VCTCop}
\end{eqnarray}
where contributions to $\VTC$ come only from separations $|\vec r-\vec r\,'| < R$ while $\VCTC$ involves only separations $|\vec r-\vec r\,'| > R$.   Note, we will distinguish the operators $\VTC$ and $\VCTC$ defined in Eqs.~\eqref{eq:VTCop} and \eqref{eq:VCTCop}, which contain a factor of $\frac{1}{2}$ and do not depend on $r$ from the functions $\VTC(r)$ and $\VCTC(r)$, defined following Eq.~\eqref{eq:truncate}, which do depend on $r$ and do not contain the factor of $\frac{1}{2}$.

As is discussed in greater detail in Sec.~\ref{sec:VTC}, the finite-range $\VTC$ term can be used directly in a lattice calculation to determine its contribution to the $\pi^+\pi^+$ scattering phase shift.  Both the QCD interactions and those implied by $\VTC$ have a finite range and can be used with Luscher's finite-volume quantization condition to determine their combined contribution to the $\pi^+\pi^+$ scattering phase shift.  We can then Taylor expand this quantization condition in $\alpha$.  The $\alpha^0$ term in this expansion gives the usual finite-volume result, determining the QCD scattering phase shift in terms of the computed finite-volume energy.  The first-order term in this expansion of Luscher's quantization condition will determine the first order contribution to that phase shift coming from $\VTC$ in terms of the first-order shift in the finite-volume energy arising from the $\VTC$ term in the QCD + QED Hamiltonian, a quantity that can be directly computed from lattice QCD as is derived in Sec.~\ref{sec:VTC}.

The contribution of $\VCTC$ to the $\pi^+\pi^+$ scattering phase shift appears inaccessible to the methods of lattice QCD because of its infinite range.  However, as a result of the minimum separation $R$ that enters $\VCTC$, the effects of $\VCTC$ come from large distances where we will show that they can be calculated analytically in terms of the QCD $\pi^+\pi^+$ scattering phase shift and the pion electromagnetic form factor if the center-of-mass energy is below the four-pion threshold.

\section{Numerical treatment of $\VTC$}
\label{sec:VTC}
Because of the spatial cutoff at the distance $R$, the truncated Coulomb potential $\VTC$ can be used in a standard finite-volume lattice QCD calculation to determine its contribution to the $\pi^+\pi^+$ scattering phase shift.  As in lattice QCD calculations that include \qedl, we can add $\VTC$ to QCD either perturbatively to some finite order in $\alpha$ or non-perturbatively including all orders in $\alpha$.  Since we wish to preserve the separation of the Coulomb interaction and the transverse radiation, it is natural to carry out this portion of the calculation to first order in $\alpha$.  We begin with L\"uscher's finite-volume quantization condition~\cite{Luscher:1990ux}:
\begin{equation}
\delta_0(p)+\phi(q) = n\pi,
\label{eq:Luscher}
\end{equation}
where for simplicity we focus on the case of practical interest, that of $s$-wave scattering.  If $E$ is the energy of a two-pion state in a finite volume with sides of length $L$ with periodic boundary conditions, then $p=\sqrt{(E/2)^2-m^2}$ must obey Eq.~\eqref{eq:Luscher} for integer $n$.  The function $\phi(q)$ obeys 
\begin{equation}
\tan\phi(q) = -\frac{\pi^{3/2}q}{\mathcal{Z}_{00}(1,q)}
\end{equation}
where
\begin{equation}
\mathcal{Z}_{00}(s,q) = \frac{1}{\sqrt{4\pi}}\sum_{\vec n \in Z^3}(\vec n\,^2-q^2)^{-s}
\end{equation}
and $q = pL/(2\pi)$.  

We then expand the quantized energy $E = E^{(0)} + \alpha E^{(1)} + \ldots$ in a power series in $\alpha$.  If we perform a similar expansion of the phase shift $\delta_0(p) = \delta_0^{(0)} (p)+ \alpha \delta_0^{(1)}(p)+\ldots$ then Eq.~\eqref{eq:Luscher} can also be expanded to relate the lattice result for $E^{(1)}$ to the desired order-$\alpha$ contribution of $\VTC$ to the scattering phase shift:
\begin{equation}
\delta_0^{(1)}(p^{(0)}) =-\left\{\frac{d \delta_0(p)}{d p} + \frac{d \phi(q)}{d q}\frac{L}{2\pi}\right\}_{p=p^{(0)}} \frac{E^{(0)}}{4p^{(0)}} E^{(1)} 
\label{eq:energy-shift}
\end{equation}
where $p^{(0)} = \sqrt{(E^{(0)}/2)^2-m^2}$.

The first-order energy $E^{(1)}$ can be determined directly from a lattice calculation in the spatial volume $L^3$.  If $O_{\pi\pi}(t)$ is a suitable $\pi\pi$ interpolating operator localized at the time $t$ which is invariant under the allowed lattice translations and rotations, then using first-order perturbation theory, $E^{(1)}$ can be determined from the ratio of correlation functions:
\begin{eqnarray}
E^{(1)} =  \frac{\left\langle O_{\pi\pi}^\dagger(t_f) \frac{1}{2}\int_V\!\!\int_V d^3r_2 d^3r_1\rho(\vec r_2,t_V)\VTC(|\vec r_2 - \vec r_1|_L)\rho(\vec r_1,t_V) O_{\pi\pi}(t_i)\right\rangle}{\left\langle O_{\pi\pi}(t_f)  O_{\pi\pi}(t_i)\right\rangle}  
\label{eq:VTC-lattice}
\end{eqnarray}
provided $t_f-t_V$ and $t_V-t_i$ are each sufficiently large that Eq.~\eqref{eq:VTC-lattice} is actually determining the matrix element of $\VTC$ between the ground state  of the $\pi^+\pi^+$ system and itself, while excited states are suppressed.  Here we have modified the argument of $\VTC$ to insure the translational symmetry supported by the periodic boundary conditions imposed on the lattice volume in which it is being used:
\begin{equation}
|\vec r_2 - \vec r_1|_L = \left\{\sum_{i=1}^3\left(\min\Bigl[\bigl|(r_2)_i-(r_1)_i\bigr|, L-\bigl|(r_2)_i-(r_1)_i\bigr|\Bigr]\right)^2\right\}^\frac{1}{2}.
\end{equation}

One additional minor complication that must be addressed in this calculation is the renormalization of the $\pi^+$ mass that results from the Coulomb interaction.  This mass shift can be computed from a three-point function very similar to that appearing in Eq.~\eqref{eq:VTC-lattice} in which the $\pi^+\pi^+$ interpolating operator $O_{\pi\pi}$ is replaced by an interpolating operator for a single pion.  This shift in mass can be eliminated by adding a first-order shift $\alpha m^{(1)}$ to the quark mass.  The quantity $m^{(1)}$ would be chosen so that when this order-$\alpha$ mass term is combined with the Coulomb potential the $\pi^+$ mass is not changed from its original value.  This first-order mass term should then be included in the calculation of $E^{(1)}$ described in Eq.~\eqref{eq:VTC-lattice} by simply adding it to the Coulomb potential operator in that equation.  The resulting value for $E^{(1)}$ could then be used in Eq.~\eqref{eq:energy-shift} to determine the first-order E\&M contribution to the scattering phase shift for two $\pi^+$ particles, each with a fixed physical mass, the same mass that was used in the original order-$\alpha^0$ QCD calculation.

\section{Analytic treatment of $\VCTC$}
\label{sec:VCTC}
In this section we will calculate the contribution of $\VCTC$ to the $\pi^+\pi^+$ scattering phase shift $\delta_l$ to first order in $\alpha$ in terms of two quantities: i) the phase shift $\delta_l$ without QED corrections and ii) the electromagnetic form factor of the pion.  This infinite-volume, order-$\alpha$ $\VCTC$ correction can be added to the correction determined numerically to first order in $\VTC$ using lattice QCD as described in the previous section to obtain the entire contribution of the instantaneous Coulomb potential to the $\pi^+\pi^+$ scattering phase shift.   This combined result should have only finite-volume errors that fall exponentially as the volume grows with the exception of the power-law corrections that come from omitting the scattering from higher partial waves in usual treatment of finite-volume quantization and whatever approximations are introduced when determining the $\pi^+$ form factor.

\subsection{General formulation}

\begin{figure}
\centering
\includegraphics[width=1.0\linewidth]{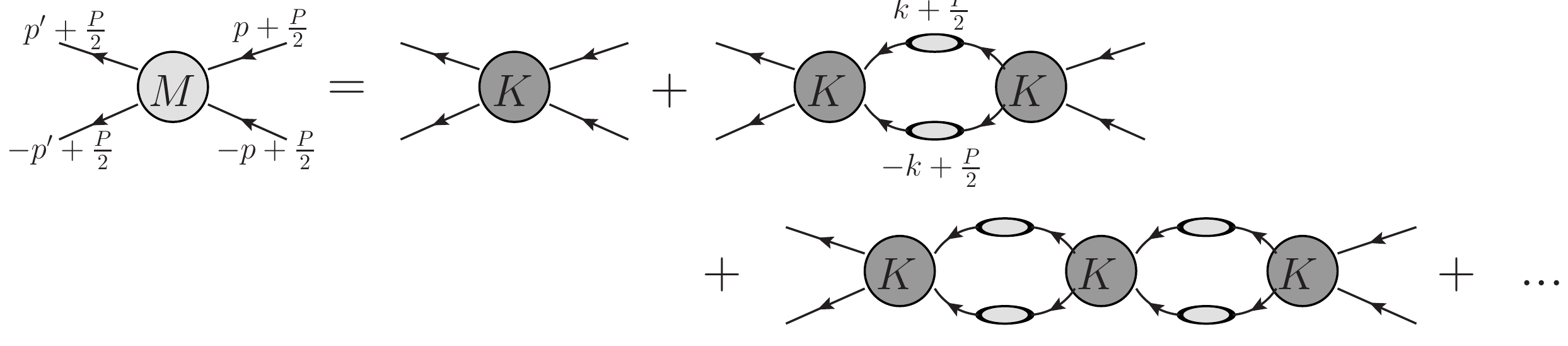}
\caption{A graphical expansion of the full scattering amplitude $M$ (with the label $M$) in terms of products factors of the two-particle irreducible kernel (with the label $K$) joined by dressed pion propagators, shown as a pion with a lightly shaded bubble representing a sum of all one-particle reducible graphs.  This is the standard expression for the complete scattering amplitude $M$ that is useful to discuss two-particle scattering below the four-particle threshold can include the effects of the Coulomb interaction $\VCTC$.}
\label{fig:LS}
\end{figure}

This analytic calculation can be performed by working with the usual relativistic Lipmann-Schwinger equation which expresses the full $\pi^+\pi^+$ scattering amplitude as a sum of products of two-particle irreducible kernels connected by pairs of pion propagators as shown in Fig.~\ref{fig:LS}.  We will refer to this sum as the Lipmann-Schwinger series.   As is conventional in such discussions we will treat the composite pion in QCD as an elementary particle in a relativistic $(\phi^\dagger\phi)^2$ field theory working to arbitrary order in a perturbation expansion in the $(\phi^\dagger\phi)^2$ interaction.  This will result in a somewhat physical discussion of the pion structure, here introduced by the $(\phi^\dagger\phi)^2$ coupling,  whose space-time scale is set by the pion mass.  As we explore the approximations necessary for the validity of our approach, we will be able to estimate their size and appreciate the bounds on the scattering energy that must be imposed.  We will then assume that the relations established are universal and will also hold true in the physical QCD problem provided that the spatial scale $R$ appearing in $\VCTC$ is larger than the distance scale of QCD so that the differences in structure between QCD and this $(\phi^\dagger\phi)^2$ model become irrelevant.  This requirement of universal behavior, independent of quark structure, implies that the truncation radius obey $R \gg \Lambda_\mathrm{QCD}$

The on-shell center-of-mass $\pi\pi$ scattering amplitude $M_l(E)$ for angular momentum $l$ that is determined by the graphs shown in Fig.~\ref{fig:LS} can be defined by
\begin{equation}
iM_l(E) \delta_{l'l} \delta_{m'm} = \frac{1}{4\pi}\int\!\!\int d\Omega_{\hat p'} d\Omega_{\hat p}\,
                                                      Y^*_{l'm'}(\hat p') M\bigl((\vec p\,',\omega_p),(-\vec p\,',\omega_p),(\vec p,\omega_p),(-\vec p,\omega_p)\bigr) Y_{lm}(\hat p).
\label{eq:PW-projection}
\end{equation}
Here the pion three-momenta $\vec p = p\hat p$ and $\vec p\,' = p\hat p\,'$ are proportional to the two unit vectors over whose directions we are integrating.  The pion energies are given by $\omega_p=+\sqrt{p^2+m^2}$ where $m$ is the pion mass and the total energy $E$ in the center-of-mass system is given by $E=2\omega_q$.  

The amplitude $M\bigl(p_4,p_3,p_2,p_1\bigr)$ can be obtained directly from the usual connected time-ordered product:
\begin{eqnarray}
&& \prod_{i=1}^4 \left\{\frac{p_i^2+m^2}{i}\right\}\prod_{i=1}^4 \int \left\{\int d^4 x_i \right\} 
e^{i(-p_4x_4-p_3x_3+p_2x_2+p_1x_1)}
                  \bigl\langle\phi(x_4)\phi(x_3)\phi^\dagger(x_2)\phi^\dagger(x_1)\bigr\rangle_\mathrm{conn} \nonumber \\
&& \hskip 1.5 in = (2\pi)^4\delta^4(p_4+p_3-p_2-p_1) M(p_4,p_3,p_2,p_1),
\label{eq:time-ordered}
\end{eqnarray}
which must be evaluated on-shell with $(p_i)_4 = \omega_p$ for $1 \le i \le 4$.  For $E < 4m$ the unitarity of the scattering matrix implies that $M_l(E)$ can be written as
\begin{equation}
M_l(E) = 32\pi\frac{\omega_p}{p}\frac{e^{2i\delta_l}-1}{2i}.
\label{eq:phase-shift}
\end{equation}
Except for a change in overall sign of the metric, these are similar to the conventions of Ref.~\cite{Christ:2015pwa}.  We will use a Minkowski metric $(1,1,1,-1)$ so that analytic continuation between Euclidean and Minkowski formulations can be accomplished by changes in phase of the energy or time arguments.

In fact, in the discussion below we will begin with a Euclidean-space version of Eq.~\eqref{eq:time-ordered} which defines the amplitude $M_E(p_4,p_3,p_2,p_1)$ for Euclidean arguments.  The amplitude $M_E$ is defined by an equation identical to Eq.~\eqref{eq:time-ordered} except that the time ordered product is to be computed using a Euclidean path integral or a Hilbert-space formalism in which the time-dependent fields are defined using exponentials of the Hamiltonian without the usual factor of $i$.  In addition, the four-vector dot products that appear in the Fourier transforms from position to momentum space in the Euclidean version of Eq.~\eqref{eq:time-ordered} use the Euclidean metric $(1,1,1,1)$.  In the discussion below we will explicitly analytically continue the amplitude $M_E$ to obtain $M$.  

In order to discuss this analytic continuation concretely, it is useful to remove the momentum-conserving delta function from Eq.~\eqref{eq:time-ordered}, to write the equation in terms of the total incoming Euclidean four-momentum $P$ and two relative four-momenta $p$ and $p'$ which are defined by the rewritten version of Eq.~\eqref{eq:time-ordered}:
\begin{equation}
M_E(p',p,P) = \int\!\!\int\!\!\int\!\! d^4 y  d^4 x_2 d^4 x_1e^{-ip'y}\left\langle\phi(\frac{y}{2})\phi(-\frac{y}{2})\phi^\dagger(x_2)\phi^\dagger(x_1)\right\rangle_\mathrm{amp} e^{i(\frac{P}{2}+p)x_2}e^{i(\frac{P}{2}-p)x_1},
\label{eq:M_E-continue}
\end{equation}
where we have used momentum conservation and translational invariance to reduce the number of four-vectors on which $M_E$ depends from four to three and to evaluate the connected four-point function at the location $x_4+x_3=0$.  In the center-of-mass system among the three four-momenta $p'$, $p$ and $P$, only $P$ has a non-zero energy component.  In the standard approach to analytic continuation from Euclidean to Minkowski space the real Euclidean energy $P_0$ is changed in phase as: $P_0 \to e^{-i\theta}P_0$ with $\theta$ increasing from zero to $\pi/2$.  In order to avoid potential exponential divergence in the integrals over $(x_1)_0$ and $(x_2)_0$, these integration variables are also rotated in phase in the opposite direction: $(x_i)_0 \to e^{i\theta}(x_i)_0$, $i=1, 2$.  This procedure then requires the time-dependent Euclidean Green's function $\left\langle\phi(\frac{y}{2})\phi(-\frac{y}{2})\phi^\dagger(x_2)\phi^\dagger(x_1)\right\rangle$ to be analytically continued in the time.   

Although this approach to analytic continuation will not be used below, this interpretation is useful to determine precisely what Minkowski amplitude results after this analytic continuation is performed.  By design the resulting Minkkowski-space Fourier transform will contain the phase whose dependence on the total four-momentum $P$ is $\frac{\vec P}{2}\cdot (\vec x_1+\vec x_2) + \frac{P_0}{2}\bigl((x_1)_0+(x_2)_0\bigr)$.  With our conventions, this is the correct phase for an incoming, positive energy only if $P_0 = -E$ where $E=2\omega_p$.  Thus, when performing this analytic continuation we will begin with $P_0=-E$ and carry out the phase rotation $P_0 = -Ee^{-i\theta}$, increasing $\theta$ from zero to $\pi/2$.

\begin{figure}
\centering
\includegraphics[width=0.6\linewidth]{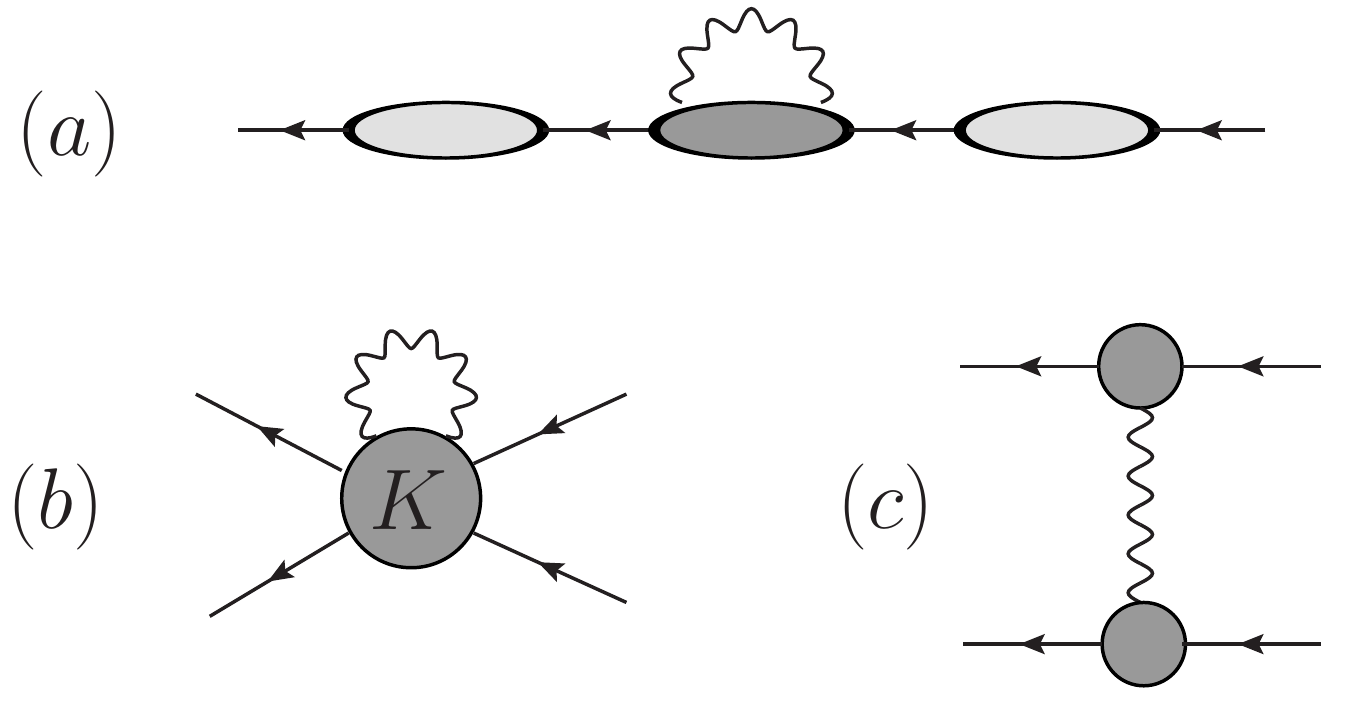}
\caption{Corrections to the subdiagrams appearing in Fig.~\ref{fig:LS} that are first-order in $\VCTC$. Diagram (a) is the correction to the dressed pion propagator expressed as two full pion propagators connected by a sum of one-particle reducible graphs (with darker shading) which cannot be separated into two parts by cutting a single pion line.  (Here two subgraphs joined by a photon line are viewed as connected.)  Diagram (b) shows the correction to the two-particle irreducible kernel which cannot be divided into two disconnected parts if the $\VCTC$ photon line is cut.  Diagram (c) represents the class of two-particle irreducible diagrams that can be divided into two disconnected parts when the $\VCTC$ photon line is cut.}
\label{fig:AddCTC}
\end{figure}

Since we wish to calculate the first order effects of $\VCTC$ on the $\pi\pi$ phase shift, we will consider those terms in which $\VCTC$ enters one of the components appearing in the Lipmann-Schwinger series shown in Fig.~\ref{fig:LS}.   The $\VCTC$ interaction can enter in three ways as illustrated Fig.~\ref{fig:AddCTC}.  In the first case $\VCTC$ contributes to the self-energy graph appearing in one of the pion propagators connecting the two-particle irreducible kernels as shown in graph (a) of Fig.~\ref{fig:AddCTC}.  The second and third cases arise when $\VCTC$ appears in the Bethe-Salpeter kernel and are distinguished by a property of the resulting kernel.  Since the $\VCTC$ interaction involves the fourth components of two electromagnetic currents at positions $(\vec z_2, t)$ and $(\vec z_1, t)$ multiplied by the function $\theta(|\vec z_2 - \vec z_1|-R)/|\vec z_2 - \vec z_1|$ we are representing this $O(\alpha)$ insertion graphically by two vertices at the positions $(\vec z_2, t)$ and $(\vec z_1, t)$ joined by a wavy line corresponding to this position-dependent function and referred to here as a ``photon'' line.  If cutting this line separates the graph in the kernel $K$ into two parts then the corresponding amplitude is of type (c).  Otherwise it is of type (b).  We will refer to these three cases as insertions of types (a), (b) and (c).  

As we will see, the $\VCTC$ contributions from graphs of type (a) and (b) are exponentially suppressed for large $R$.  This will be demonstrated in the next section by an analytic continuation from Euclidean space in which the evaluations of the self-energy and Bethe-Salpeter kernel graphs remain in Eulidean space where the exponential suppression coming from the large space-like separation of the two vertices in $\VCTC$ can be easily seen.  Diagrams of type (c), which contain the Coulomb scattering of two far-separated pions, will therefore contain all of the relevant effects of $\VCTC$ and are evaluated in Section~\ref{sec:LD_scattering}.

\subsection{Analytic continuation}
\label{sec:AC}

In order to show that the contribution $\VCTC$ to graphs of type (a) and (b) is exponentially suppressed for large $R$ for the case of $\pi^+\pi^+$ scattering at an energy below the four-pion threshold, we will first demonstrate that these self-energy and kernel subgraphs can be evaluated in Euclidean space even after the original Euclidean amplitude $M_E$ has been analytically continued to physical Minkowski energies.  To show this, we propose a particular procedure to carry out this analytical continuation in which the external lines and the internal two-pion loop integrals shown explicitly in Fig.~\ref{fig:LS} are evaluated in momentum space and the loop integration contours distorted to avoid the singularities as this continuation  is carried out.  However, during this process the self-energy and kernel graphs will be evaluated as functions of position and those positions (and the amplitudes in which they appear) will remain in Euclidean space.

Anticipating this strategy we express both the self-energy and two-particle irreducible kernels as Fourier transforms of Euclidean, position-space functions:
\begin{eqnarray}
D(P,k)  &=& \int d^4 x \bigl\langle\phi(0)\phi^\dagger(x)\bigr\rangle_{\mathrm{1PI}} e^{i(k+\frac{P}{2})x} 
\label{eq:1PI_self-energy} \\
K(P,k',k) &=& \int\!\!\int\!\!\int d^4 y\, d^4 x_2\, d^4 x_1
\label{eq:2PI_kernel} \\
              && \hskip 0.75 in e^{-ik'y} \left\langle \phi(\frac{y}{2})\phi(-\frac{y}{2})\phi^\dagger(x_2)\phi^\dagger(x_1)\right\rangle_{\mathrm{2PI,amp}} e^{i(k+\frac{P}{2})x_2} e^{i(-k+\frac{P}{2})x_1} .
\nonumber
\end{eqnarray}
Both expressions are to be evaluated in Euclidean space and Fourier transformed with Euclidean four-momenta.  Note we have used the translational invariance of both types of diagram to remove one of the space-time arguments from each Green's function.  

The Green's functions appearing in Eqs.~\eqref{eq:1PI_self-energy} and \eqref{eq:2PI_kernel} are intended to be entirely general including both arbitrary orders in the $(\phi^\dagger\phi)^2$ interaction and zeroth or first order in $\VCTC$.  While $\VCTC$ results from electromagnetism, since it contains only operators that are evaluated at the same time, it does not change the structure of the intermediate states that determine the time dependence of these Green's functions.  As a result, the possible presence of $\VCTC$ in these Green's functions will be ignored when discussing their Euclidean time dependence.

\begin{figure}
\centering
\includegraphics[width=0.7\linewidth]{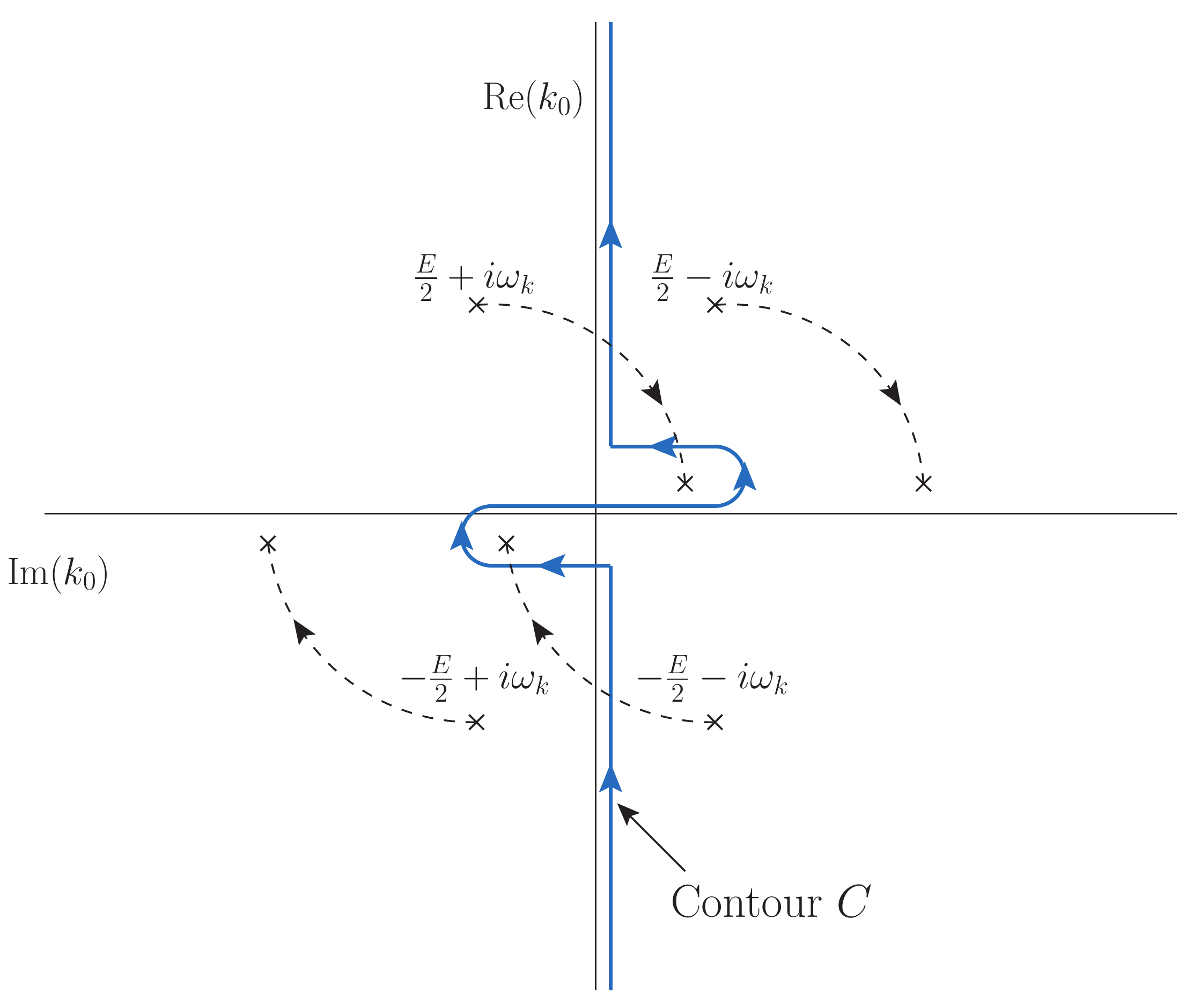}
\caption{Diagram showing the treatment of the contour of $k_0$ integration needed in the two-pion loop integrals that appear explicitly in the Lipmann-Schwinger series, using the momentum assignment shown in Fig.~\ref{fig:LS}.  The change in the incoming Euclidan energy from $E \to -i|E|$ causes the distortion of the $k_0$ contour from the vertical axis to the new contour $C$.  This diagram refers to Euclidean-space coordinates but follows an unconventional assignment of directions for the real and imaginary $k_0$ axes to give it a more familiar, Minkowski-space orientation.}
\label{fig:momentum_routing_E}
\end{figure}

With this approach, it is straight-forward to perform the needed analytic continuation in the total incoming energy $E$ from its Euclidean to its Minkowski value: replace $E$ by $e^{-i\theta}E$ and vary $\theta$ from zero to $\pi/2$.  This complex variable $E$ enters the amplitudes appearing in the Lipmann-Schwinger series in two places.  First, it appears in the exponents written explicitly in Eqs.~\eqref{eq:1PI_self-energy} and \eqref{eq:2PI_kernel}.  Since the exponential function is analytic, the continuation can be easily performed.  However, when $E$ becomes imaginary, the Fourier integrals acquire an exponentially growing factor and we must demonstrate these integrals remain convergent.   

Second, $E$ appears in the internal propagators shown in Fig.~\ref{fig:LS}.  Here the variation in $E$ from a real to an imaginary value will force the integration contour over the loop variable $k_0$ to be distorted to avoid the singularities which move as the phase of $E$ changes.  The motion of these singularities is shown by the dotted lines in Fig.~\ref{fig:LS}.  This distortion of the contours of the loop energies will require a further analytic continuation of the kernel and self-energy functions which can also be carried out since these energy variables also appear explicitly in analytic exponential functions.  Of course, we must demonstrate that this further introduction of exponentially growing time dependence does not cause these internal Fourier transforms to diverge.

Thus, our demonstration that the contribution of $\VCTC$ when appearing in diagrams of type (a) and (b) is exponentially suppressed as $R$ increases proceeds in two steps.  First we will show that the analytic continuation described above is possible, allowing the actual evaluations of the quantities $D$ and $K$ to be performed in Euclidean space.  Second we examine the case when $\VCTC$ appears in either of these Euclidean-space quantities and use their Euclidean-space character to show this exponential suppression as $R$ increases.

To address the first step we examine the convergence of the integrals over time in Eqs.~\eqref{eq:1PI_self-energy} and \eqref{eq:2PI_kernel} when the Euclidean energy appearing in the time-dependent exponent has become imaginary.  For the self-energy case described by Eq.~\eqref{eq:1PI_self-energy}, we can see from Fig~\ref{fig:momentum_routing_E} that the magnitude of the imaginary part of $k_0$ will be non-zero when $E/2>\omega_k$ and can be limited to $E/2-\omega_k<E/2-m$.  Thus, the largest exponential growth than can result from the continued Fourier transform shown in Eq.~\eqref{eq:1PI_self-energy} with the exponent $\frac{P_0}{2}+k_0$ takes the form $\exp\bigl\{(E-m)x_0\bigr\}$.  However, the Euclidean-space, one-particle-irreducible, self-energy graphs which enter Eq.~\eqref{eq:1PI_self-energy} themselves decrease exponentially as the time separation $x_0$ increases.  This decreasing behavior will be determined by the least massive intermediate state that can appear between the two interpolating operators $\phi(0)$ and $\phi^\dagger(x)$ in Eq.~\eqref{eq:1PI_self-energy} which must be a three-pion state with energy no smaller than $3m$.  Thus, our analytic continuation strategy will fail when $E-4m > 0$, reminding us of the known requirement that we stay below the four-pion threshold in our study of physical $\pi\pi$ scattering.  

The behavior of the analytically continued expression given in Eq.~\eqref{eq:2PI_kernel} for the two-particle irreducible kernel is very similar.  Now there are three time integrals which may diverge as the time separations among the arguments in Eq.~\eqref{eq:2PI_kernel} grow.  In examining the behavior expected, we might distinguish the case where this kernel is the first or last factor in the Lipmann-Schwinger series from the case where it appears somewhere in the middle.  Since we are working in the center-of-mass system, the variables $k_0$ or $k'_0$ would be zero in the first case while they can have an imaginary part as large as $E/2-m$ in the second.  Thus, we will consider the second case since it includes the first.

\begin{figure}
\centering
\includegraphics[width=0.9\linewidth]{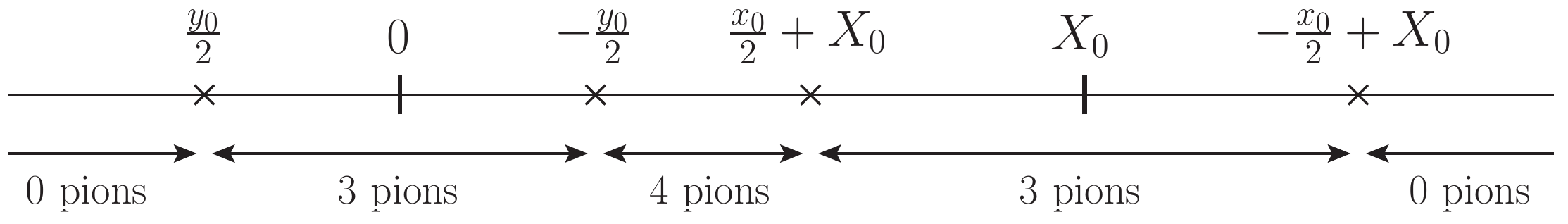}
\caption{Diagram showing the ordering of arguments used in Eq.~\eqref{eq:K-analyticity} with time increasing from right to left.  The average time of the two incoming vertices is $X_0$ which is chosen negative while the earliest outgoing vertex (at the time $-y_0/2$) is assumed to be later than the latest incoming vertex (at the time $x_0/2+X_0$).  The times of the four vertices are marked by crosses while the average incoming and average outgoing times are each shown as vertical lines.  The minimum number of intermediate pions which appear in the amplitude $K(P,k',k)$ in each interval is also indicated.}
\label{fig:ordered-line}
\end{figure}

It is convenient to replace the variables $x_2$ and $x_1$ in Eq.~\eqref{eq:2PI_kernel} by the average and relative coordinates $X=(x_2+x_1)/2$ and $x = (x_2-x_1)$.  Given the symmetry between the values $\pm y_0$ and between the values $\pm x_0$ we will chose both $x_0$ and $y_0$ to be positive.  Further we will first consider the case where $X_0$ is negative and its magnitude is larger than $(x_0+y_0)/2$.  This case is illustrated in Fig.~\eqref{fig:ordered-line}.  Because we are discussing the dependence of the two-particle irreducible kernal on its four time coordinates, such a one-dimensional figure is sufficient for a general discussion.  An amputated pion line joins the two-particle-irreducible kernel $K$ at each of the four times, $\pm \frac{y_0}{2}$ and $\pm \frac{x_0}{2}+X_0$ which implies that at each of these times the number of pions much change by an odd number.  If we include the fact the kernel must be both one- and two-particle irreducible, the allowed minimum number of pions appearing in the three intermediate states between these four times must be those shown in Fig.~\eqref{fig:ordered-line}.  

We conclude that the dependence of the integrand on the three times $x_0$, $y_0$ and $X_0$ is given by 
\begin{equation}
\left[e^{y_0(\frac{E}{2}-m)} e^{x_0(\frac{E}{2}-m)}e^{-X_0 E}\right]
\left[ e^{-3m|y_0|} e^{-4m|-\frac{y_0}{2}-(\frac{x_0}{2}+X_0)|} e^{-3m|x_0|}\right] =
e^{\frac{x_0+y_0}{2}(E-4m)} e^{-X_0(E-4m)}.
\label{eq:K-analyticity}
\end{equation}
The first three factors on the left-hand side come from the exponents in the Fourier transform appearing in Eq.~\eqref{eq:2PI_kernel} while the remaining three factors come from inserting three-pion states between $\phi(\frac{y_0}{2})$ and $\phi(-\frac{y_0}{2})$ and between $\phi^\dagger(\frac{x_0}{2}+X_0)$ and $\phi^\dagger(-\frac{x_0}{2}+X_0)$ while a four-pion intermediate state must be inserted between $\phi(-\frac{y_0}{2})$ and $\phi^\dagger(\frac{x_0}{2}+X_0)$. 

Inspecting the right-hand side of Eq.~\eqref{eq:K-analyticity}, we recognize that the integrand decreases if any of the three positive quantities $y_0$, $x_0$ and $-X_0$ is increased provided E is below the four-pion threshold, $E<4m$.
If we treat $y_0$ and $x_0$ as fixed can then discuss the behavior of the integrand as $X_0$ increases from it assumed large negative value.  As $X_0$ increases we reach the point where the lower end of the interval $[\frac{y_0}{2},-\frac{y_0}{2}]$ collides with the upper end of the interval $[\frac{x_0}{2}+X_0,-\frac{x_0}{2}+X_0]$.  At this point the $X_0$ behavior of the integrand changes from $\exp\{-X_0(E-4m)\}$ to $\exp\{-X_0(E-2m)\}$.  Thus, the integrand now decreases as $X_0$ becomes more positive.  Continuing to increase $X_0$ the next change in behavior occurs when one of the two intervals $[\frac{y_0}{2},-\frac{y_0}{2}]$ and $[\frac{x_0}{2}+X_0,-\frac{x_0}{2}+X_0]$ lies within the other and the $X_0$ behavior becomes more rapidly decreasing $\exp\{-X_0E\}$.  Next as $X_0$ increases further and these two intervals partially overlap and then separate the exponential decrease becomes steeper changing to $\exp\{-X_0(E+2m)\}$ and finally $\exp\{-X_0(E+4m)\}$.  Thus the maximum is reached when $-\frac{y_0}{2} = \frac{x_0}{2} +X_0$ at which point the integrand behaves as $\exp\{(y_0+x_0)(E-4m)\}$ insuring convergence for $E< 4m$ and providing the analyticity needed for our explicit Euclidean-space evaluation at physical energies below the four-pion threshold.  The case where $-\frac{y_0}{2} = \frac{x_0}{2} + X_0$ could have been anticipated as giving the largest contribution since it avoids the situation with a four-pion intermediate state, {\it e.g.} in Fig.~\ref{fig:ordered-line} the 
case where the ``4 pions'' segment has zero length.

We conclude that the Minkowski-space $\pi\pi$ scattering amplitude can be determined from the Lipmann-Schwinger series in which the self-energy corrections and two-particle irreducible kernel are computed in Euclidean space.  This implies that when $\VCTC$ appears in those subgraphs, its contribution will be suppressed exponentially when $R$ becomes large.  In each case, the two vertices introduced by $\VCTC$, separated by the distance $R$, must be joined by at least two Euclidean pion propagators so that the $\VCTC$ contribution will decrease exponentially as $R$ grows -- the property that we wish to establish.  

\begin{figure}
\centering
\includegraphics[width=0.6\linewidth]{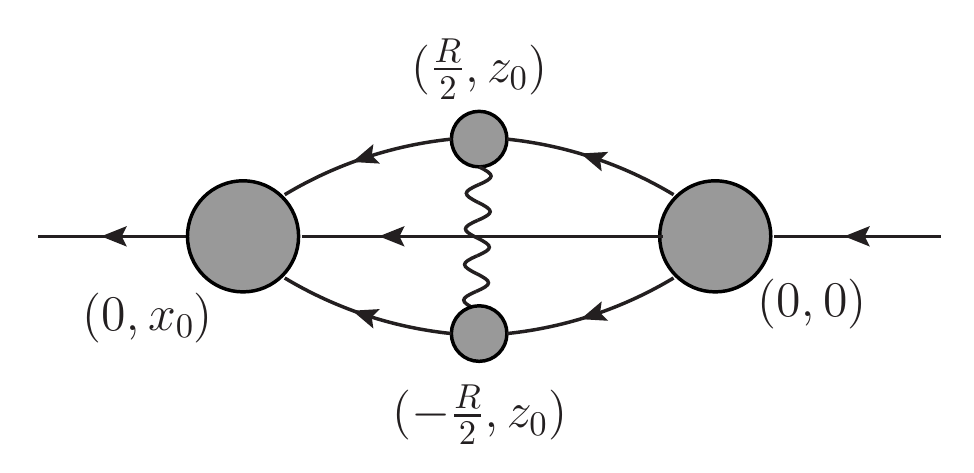}
\caption{An example of the insertion of the potential $\VCTC$ into a self-energy graph that can be used to estimate the rate of exponential decrease of such an amplitude as the truncation radius $R$ increases.  The location in space and time of each of the four vertices is shown.  As is discussed in the text the contribution of graphs with this topology in the Lipman-Schwinger series will decrease for large $R$ as $\exp\{-\sqrt{E(4m-E)}R\}$.}
\label{fig:diamond}
\end{figure}

We should point out that an estimate of this exponential decrease given by $\exp\{-2mR\}$ coming simply from two sequences of single pion propagtors joining one $\VCTC$ vertex to the other is an overestimate of the rate of decrease because of the exponential growth of the Fourier transform factors discussed above.  For example, once the large spatial separation $R$ has been introduced into the self-energy or kernel subdiagrams from $\VCTC$ there will be an advantage to a diamond pattern, such as that shown Fig.~\ref{fig:diamond} for the case of a self-energy subdiagram, in which the two spatially-separated vertices in $\VCTC$ are joined to two temporally-separated vertices.  With this arrangement the exponential suppression associated with the time separation of the two additional vertices is partially compensated by the exponentially growing Fourier factors.   For the case shown in Fig.~\ref{fig:diamond}, such a diamond geometry leads to a weaken exponential suppression.  The total exponent corresponding to the four pion propagators shown in Fig.~\ref{fig:diamond} can be minimized by the choice of $z_0= x_0/2$ and a value of $x_0$ which grows as $E$ approaches $4m$ giving a reduced exponential fall off $\propto \exp\{-R\sqrt{E(4m-E)}\}$ which varies from the expected $\exp\{-2mR\}$ when $E=2m$ to zero as $E$ increases to the four-pion threshold.

\subsection{Lipman-Schwinger equation solution to first order in $\VCTC$}
\label{sec:LD_scattering}

\begin{figure}
\centering
\includegraphics[width=0.8\linewidth]{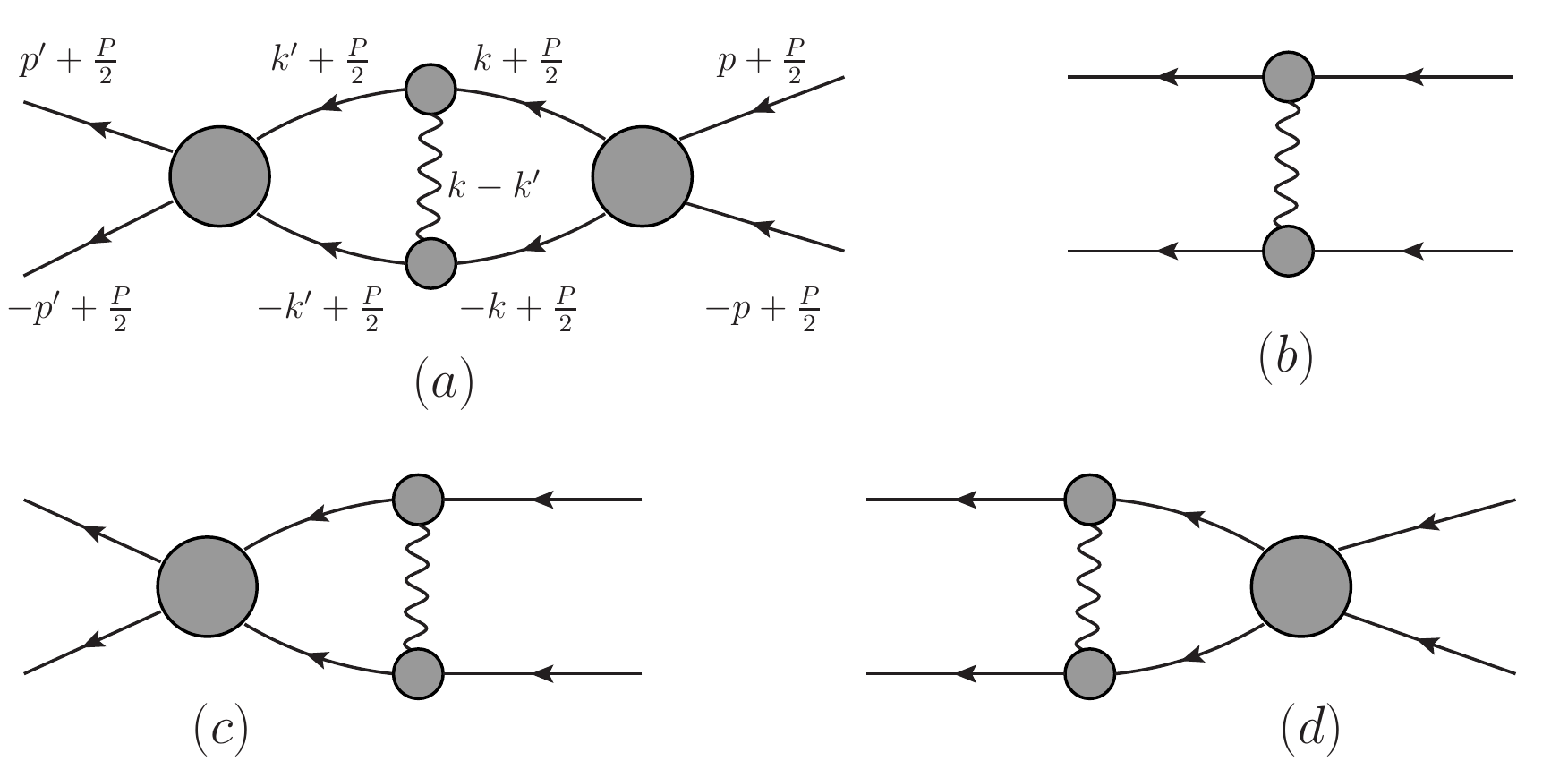}
\caption{The four types of diagram that must be evaluated to determine the correction to the scattering phase shift arsing from $\VCTC$.  Here the pion self-energy insertions on the internal lines are not shown.  In this Section a self-energy insertion carrying four-momentum $k$ is written as the product of a free-particle propagator and a function of $k^2$ which is absorbed into the off-shell scattering amplitude $M$.}
\label{fig:DWBA}
\end{figure}

The result of the previous section shows that the contribution to $\pi^+\pi^+$ scattering from $\VCTC$ will come only from diagrams of type (c) in Fig.~\ref{fig:AddCTC} once terms that are exponentially suppressed for large $R$ have been neglected.  In this section we will evaluate that contribution, in this same, exponentially-accurate approximation.  This calculation receives contributions from the four diagrams shown in Fig.~\ref{fig:DWBA}.   While the diagrams shown in that figure involve the off-shell Bethe-Salpeter scattering kernel and off-shell pion electromagnetic vertex, the presence of $\VCTC$ will be shown to restrict their evaluation to an on-shell, long-distance region allowing them to be evaluated directly in terms of infinite-volume scattering data and the on-shell pion form factor, both quantities that can be directly determined from lattice QCD or, if we choose, from a combination of experiment and dispersion theory/phenomenology.  The result will be an explicit formula determining the contribution of $\VCTC$ to the $\pi^+\pi^+$ scattering phase shift.  This is the missing component that was omitted from the phase shift calculation using finite-volume lattice methods which include only $\VTC$.   We note that all of the variables and formulae in this and later sections are expressed using Minkowski space conventions.

This calculation can be cast into a suggestive form by expressing the first-order contribution of $\VCTC$ to $M_l$ of Eq.~\eqref{eq:PW-projection} given by the sum of the four classes of diagram shown in Fig.~\ref{fig:DWBA} as a product of incoming and outgoing factors multiplied by the $\VCTC$ scattering kernel:
\begin{equation}
{\MCTC}_{,\,l} =\frac{1}{2l+1}\sum_{m=-l}^l \int \!\! \int d^4k' d^4k \, \Psi^\mathrm{out}_{lm}(k',P)^* \KCTC(k',k,P)\Psi^\mathrm{in}_{lm}(k,P). 
\label{eq:DWBA}
\end{equation}
Here the $\VCTC$ scattering kernel $\KCTC(k',k,P)$ is shown as diagram (c) in Fig.~\ref{fig:AddCTC} and can be written in momentum space after the total momentum conserving delta function has been removed:
\begin{eqnarray}
\KCTC(k',k,P) &=& -2\int\!\! \int d^4 y_2\, d^4 y_1 \int d^3 z \int \!\! \int d^4 x_2\, d^4 x_1 \;
e^{-i(k'+\frac{P}{2})y_2}e^{-i(-k'+\frac{P}{2})y_1} 
\label{eq:KCTC-MS} \\ 
&& \hskip 0.1 in
   \Bigl\langle \phi(y_2)\rho(\frac{z}{2})\phi^\dagger(x_2)\Bigr\rangle_{\mathrm{1PI}}   \VCTC(z)
   \Bigl\langle \phi(y_1)\rho(-\frac{z}{2})\phi^\dagger(x_1)\Bigr\rangle_{\mathrm{1PI}} e^{+i(k+\frac{P}{2})x_2}e^{i(-k+\frac{P}{2})x_1}. 
   \nonumber
\end{eqnarray}
The factor of two results from a combination of the four equivalent contractions contributing to $\KCTC(k',k,P)$ and the factor of $\frac{1}{2}$ appearing in the operator $\VCTC$ defined in Eq.~\eqref{eq:VCTCop}.  This expression is analogous to that appearing in Eq.~\eqref{eq:2PI_kernel} for the complete scattering kernel except we have used translation symmetry to fix the origin at the midpoint between the locations of the two charge density operators.

As suggested by Fig.~\ref{fig:DWBA}, $\Psi^\mathrm{in/out}_{lm}(k,P)$ can be separated into two components:
\begin{equation}
\Psi^\mathrm{in/out}_{lm}(k,P) = \psi^0_{lm}(k,P) + \psi^\mathrm{in/out}_{lm}(k,P).
\label{eq:wave-function-dcomp}
\end{equation}
The first component, $\psi^0_{lm}(x,P)$, is constructed from the plane-wave component of the incoming state which corresponds to no initial/final state scattering beyond that resulting from $\VCTC$.  The second component includes all initial/final scattering and involves the full off-shell scattering amplitude $M(E)$ defined in Eq.~\eqref{eq:time-ordered}.  We will now consider in turn the calculation of the contribution from $\psi^0_{lm}(k,P)$ and then $\psi^\mathrm{in/out}_{lm}(k,P)$.

\subsubsection{Plane-wave, $\psi^0_{lm}(k,P)$, contribution}
\label{sec:psi_0}

The plane wave part, $\psi^0_{lm}(k,P)$ can be deduced directly from Eqs.~\eqref{eq:PW-projection},  \eqref{eq:time-ordered} and \eqref{eq:M_E-continue}:
\begin{eqnarray}
\psi^0_{lm}(k,P) &=& \frac{1}{(2\pi)^4}\int d^4 x\; e^{-i\vec k\cdot \vec x} \left\{\frac{1}{\sqrt{4\pi}}\int d\Omega_{\hat p} e^{i\vec p\cdot \vec x} Y_{lm}(\hat p) \right\} \\
                    &=&\frac{1}{\sqrt{4\pi}p^2}\delta(|\vec k|-p)\delta(k_0)Y_{lm}(\hat k).
\label{eq:plane-wave}
\end{eqnarray}
Here $p=\sqrt{(E/2)^2-m^2}$ and $\vec p = p \hat p$.   Although $\psi^0_{lm}(k,P)$ can be easily determined as in Eq.~\eqref{eq:plane-wave}, when we consider the second term, $\psi^\mathrm{in/out}_{lm}(k,P)$, it will be simplest to directly evaluate its contribution to the product $\left[\KCTC\Psi^{\mathrm{in/out}}\right](k',P)$.  If this is done for $\psi^0_{lm}(k,P)$, we find:
\begin{eqnarray}
\left[\KCTC\psi^0_{lm}\right](k',P) &=& -2\int d^3 z\int d^3 k\Biggl\{ \Gamma_0\Bigl(k'+\frac{P}{2},\bigl(\vec k,\frac{E}{2}\bigr)\Bigr) e^{i(\vec k - \vec k')\cdot\vec z} \VCTC\bigl(|\vec z|\bigr)  
\label{eq:K-psi0-1} \\
&& \hskip 1.2 in 
\cdot \Gamma_0\Bigl(-k'+\frac{P}{2},\bigl(-\vec k, \frac{E}{2}\bigr)\Bigr)\frac{1}{\sqrt{4\pi} p^2}
\delta(|\vec k|-p)Y_{lm}(\hat k) \Biggr\} \nonumber \\
&=& -\frac{1}{\sqrt{\pi}}\int d^3 z\int d\Omega_{\hat p}\; \Gamma_0\Bigl(k'+\frac{P}{2},\bigl(\vec p, \frac{E}{2}\bigr)\Bigr) e^{i(\vec p - \vec k')\cdot\vec z} \VCTC\bigl(|\vec z|\bigr)  \label{eq: psi-0}\\
&& \hskip 1.2 in 
\cdot \Gamma_0\Bigl(-k'+\frac{P}{2},\bigl(-\vec p, \frac{E}{2}\bigr)\Bigr) Y_{lm}(\hat p).
\nonumber
\end{eqnarray}
As in previous equations the two arguments of the pion vertex function $\Gamma_0(q',q)$ are four-vectors but here the incoming four-momenta are written out as an explicit combination $(\vec q, q_0)$ of spatial, $\vec q$, and temporal, $q_0$, components.

An important simplifying step in the present discussion and that below considering $\psi^\mathrm{in/out}_{lm}(k,P)$ involves the treatment of the vertex functions $\Gamma_0(\frac{P}{2}\pm k',\frac{P}{2} \pm k)$.  For both of these functions to be on-shell, four conditions must be obeyed: 
\begin{eqnarray}
k_0=k'_0=0 \quad\quad |\vec k|=|\vec k\,'|=\sqrt{\left(\frac{E}{2}\right)^2-m^2}.
\label{eq:on-shell}
\end{eqnarray}
If these conditions are obeyed then each vertex function can be expressed in terms of a single covariant form factor $F$ depending only on the momentum transfer $(k'-k)^2= |\vec k -\vec k\,'|^2$:
\begin{equation}
\Gamma_0\left(\pm k'+\frac{P}{2},\pm k+\frac{P}{2}\right) = E F\bigl((\vec k'-\vec k)^2 \bigr).
\label{eq:FormFactor}
\end{equation}
We can then introduce a charge density function $\overline{\rho}(r)$ given by:
\begin{equation}
 \overline{\rho}(|\vec r|) = \frac{1}{(2\pi)^3} \int d^3 q e^{i\vec q \cdot \vec r} F(\vec q\,^2).
\label{eq:charge-density}
\end{equation}
Thus, if Eq.~\eqref{eq:on-shell} is obeyed we can write
\begin{equation}
\Gamma_0\left(\pm k'+\frac{P}{2},\pm k+\frac{P}{2}\right) = E \int d^3 r\, e^{\pm(\vec k - \vec k\,')\cdot \vec r} \overline{\rho}(r).
\label{eq:FormFactor-PS}
\end{equation}
Here the extra bar has been introduced to distinguish the real function $\overline{\rho}(r)$ from the operator $\rho(r)$ introduced in Section~\ref{sec:formulation}.  

These on-shell conditions are obeyed for the case that $\psi^0$ appears in both the right and left factors.  As we will see below for the case of $\psi^\mathrm{in/out}$, in the limit of large $R$ the effect of the two single-pion poles in the integrand of the integrals over the four-vectors $k$ and $k'$ in Eq.~\eqref{eq:DWBA} is also to limit the contributing values of $k$ and $k'$ to those obeying the conditions in Eq.~\eqref{eq:on-shell}.  Thus, in the case where $k$ and $k'$ are off-shell we will view Eq.~\eqref{eq:FormFactor-PS} as an approximation where terms that vanish on-shell have been dropped.  Since those unwanted terms will come with factors which cancel the poles which contribute to the large $R$ limit, the neglect of these terms is justified, as we will see, if terms that fall exponentially for large $R$ are neglected.

For the case at hand, if Eq.~\eqref{eq:FormFactor-PS} is used to replace each of the two vertex functions in Eq.~\eqref{eq: psi-0}, that equation can be rewritten as:
\begin{eqnarray}
\left[\KCTC\psi^0_{lm}\right](k',P) &=&-\frac{E^2}{\sqrt{\pi}}\int\!\!\int\!\!\int\!\!\int d^3 w\, d\Omega_{\hat p}\,d^3 r_2\, d^3 r_1 \label{eq:psi0-result} \\
&& \hskip 1.0 in \overline{\rho}\bigl(|\vec r_2|\bigr) \VCTC\bigl(|\vec w -\vec r_2 +\vec r_1|\bigr) \overline{\rho}\bigl(|\vec r_1|\bigr) 
e^{i\vec w \cdot (\vec p - \vec k\,')} Y_{lm}(\hat p), \nonumber
\end{eqnarray}
provided we assume that $k'$ can be treated as on-shell and substitute $\vec z = \vec w - \vec r_2 +\vec r_1$

This result for the plane-wave piece $\psi^0$ takes a recognizable form if we examine the case where the plane-wave factor $\psi^0$ also appears as the left-hand factor and use the standard decomposition of the plane wave $e^{i\vec w\cdot(\vec p - \vec p\,')}$ in spherical harmonics and spherical Bessel functions.  If we define ${\MCTC^{00}}_{,l}$ as the resulting purely plane-wave contribution to ${\MCTC}_{,l}$ then following the conventions of Ref.~\cite{Jackson:1998nia} we find
\begin{eqnarray}
{\MCTC^{00}}_{,l} &=& \frac{1}{2l+1}\sum_{m=-l}^l\int\!\!\int d^4k'\,d^4 k\,{\psi^0_{lm}(k',P)}^* \KCTC(k',k,P) \psi^0_{lm}(k,P) \\
                       &=& -2E^2 \int\!\!\int\!\!\int d^3w\, d^3 r_2\, d^3 r_1\, j_l^2(p w)\label{eq:BA-1} 
\\
   &&\hskip 1.0 in \cdot \overline{\rho}\bigl(|\vec r_2|\bigr) \VCTC\bigl(|\vec w -\vec r_2 +\vec r_1|\bigr) \overline{\rho}\bigl(|\vec r_1|\bigr)
 \nonumber \\
            &=& 32\pi\frac{\omega_p}{p}\dCTCZ_l, \label{eq:BA-2}
\end{eqnarray}
where to obtain Eq.~\eqref{eq:BA-2} we have expanded Eq.~\eqref{eq:phase-shift} to first-order in $\alpha$ to relate the expression computed in Eq.~\eqref{eq:BA-1} to the scattering phase shift $\dCTCZ_l$ caused by $\VCTC$ alone, assuming no strong interaction scattering (but including the effects of the pion form factor).  If Eq.~\eqref{eq:BA-2} is solved for $\dCTCZ_l$ we obtain
\begin{equation}
\dCTCZ_l = -\frac{p\,\omega_p}{4\pi}\int\!\!\int\!\!\int d^3w\, d^3 r_2\, d^3 r_1\, j_l^2(p w) 
 \overline{\rho}\bigl(|\vec r_2|\bigr) \VCTC\bigl(|\vec w -\vec r_2 +\vec r_1|\bigr) \overline{\rho}\bigl(|\vec r_1|\bigr).
\label{eq:BA-3}
\end{equation}
the standard relativistic Born approximation for the scattering of two identical mesons, each with a charge distribution $\overline{\rho}(|\vec r|)$, by the potential $\VCTC$.

\subsubsection{Interacting $\psi^\mathrm{in/out}_{lm}(k,P)$ contribution}
\label{sec:psi_in-out}

The second scattering term $\psi^\mathrm{in/out}_{lm}(k,P)$ in Eq.~\eqref{eq:wave-function-dcomp} takes a similarly simple form for the case of $\KCTC$.  We will examine the case of  $\psi^\mathrm{in}_{lm}(k,P)$, starting with the explicit definition of the Feynman amplitude implied by the left-hand factor in diagrams (a) or (d) of Fig.~\ref{fig:DWBA}:
\begin{eqnarray}
\psi^\mathrm{in}_{lm}(k,P) = -\frac{1}{\sqrt{4\pi}(2\pi)^4} 
\frac{1}{\bigl[(\frac{P}{2}+k)^2+m^2\bigr]\bigl[(\frac{P}{2}-k)^2+m^2\bigr]} \int d\Omega_{\hat p}M(k,p,P) Y_{lm}(\hat p),
\label{eq:psi-in_1}
\end{eqnarray}
where we begin with the four-momentum $k$ and the scattering amplitude $M_E(k,p,P)$, defined in Eq.~\eqref{eq:M_E-continue}, in Euclidean space.  As in the previous examination of $\psi^0_{lm}(k,P)$ we study the combination of the kernel $\KCTC(k',k,P)$ with $\psi^\mathrm{in}_{lm}(k,P)$
\begin{eqnarray}
\left[\KCTC\psi^\mathrm{in}_{lm}\right](k',P) &=& \int\!\!\!\int d^3 z d^4 k\;\Gamma_0\Bigl(k'+\frac{P}{2},k+\frac{P}{2}\Bigr) e^{i(\vec k - \vec k')\cdot\vec z} \VCTC\bigl(|\vec z|\bigr)
                        \Gamma_0\Bigl(-k'+\frac{P}{2},-k+\frac{P}{2}\Bigr)   \nonumber \\
&& \hskip -0.2 in 
\cdot\frac{1}{\sqrt{4\pi}(2\pi)^4}\frac{1}{\bigl[(\frac{P}{2}+k)^2+m^2\bigr]\bigl[(\frac{P}{2}-k)^2+m^2\bigr]} \int d\Omega_{\hat p}M(k,p,P) Y_{lm}(\hat p).  
\label{eq:psi-in_2}
\end{eqnarray}
Note that the combinatoric factor of two that appeared as a prefactor in Eq.~\eqref{eq:K-psi0-1} is absent here because of an additional factor of $\frac{1}{2}$ that must be introduced when $K$ and $M$ are combined.  Since both $K$ and $M$ are defined as amputated Green's functions, they each contain a combinatoric factor of two that will appear only  once for the contractions joining the $K$ and $M$ subdiagrams.  The factor of $\frac{1}{2}$ removes this extra factor.

\begin{figure}
\centering
\includegraphics[width=0.7\linewidth]{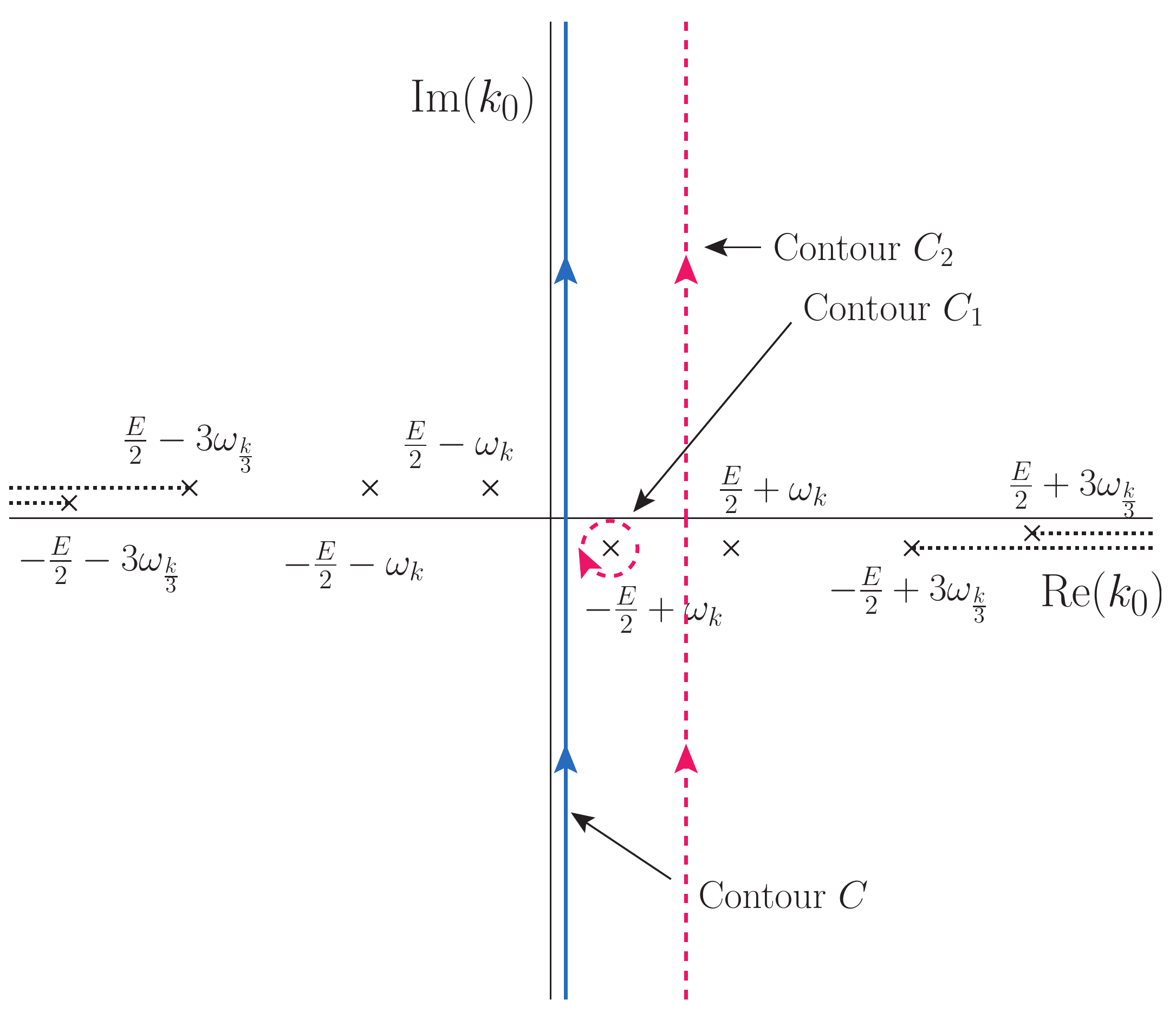}
\caption{Diagram showing the treatment of the $k_0$ contour of integration when evaluating the first-order contribution of $\VCTC$ in Eq.~\eqref{eq:psi-in_1}.  This is a copy of Fig.~\ref{fig:momentum_routing_E} but with the integration variable $k_0$ shown as a Minkowski rather than a Euclidean energy.  This change is accomplished by relabeling the axes used in Fig.~\ref{fig:momentum_routing_E}.  The familiar two-particle threshold occurs when as $E$ increases, the pole at $\frac{E}{2}-\omega_k$ collides with the pole at $-\frac{E}{2}+\omega_k$ and when $k=0$ at the endpoint of the integral over $|\vec k|$.  In this diagram we also show the location in the $k_0$ plane of the three-particle branch cuts that arise from the $k_0$ dependence of the self-energy and two-particle irreducible kernel.  These are shown as horizontal dotted lines extending to the right from the points $\pm \frac{E}{2} + 3\omega_\frac{k}{3}$ and to the left from the points $\pm \frac{E}{2} - 3\omega_\frac{k}{3}$,  moved below or above the real line for clarity.}
\label{fig:momentum_routing_M}
\end{figure}

The next step is to perform the integral over $k_0$ along the contour $C$ shown in Fig.~\ref{fig:momentum_routing_M}.  Here we have reproduced Fig.~\ref{fig:momentum_routing_E} but adopted the more familiar Minkowski labeling where the abscissa is the real part of $k_0$ which is now a standard Minkowski energy while the ordinate is the imaginary part of $k_0$.   We have also shown the location of the $k_0$ branch points corresponding to the three-pion state with total spatial momentum $\vec k$.

We will simplify the integral over $k_0$ in Eq.~\eqref{eq:psi-in_2} by using Cauchy's theorem to move the contour in that figure to the right.  As shown in Fig.~\ref{fig:momentum_routing_M} the new contour has two parts.  The first is a closed, circular clockwise contour $C_1$ surrounding the pole at $k_0=\omega_k-\frac{E}{2}$.  The second piece is a vertical contour $C_2$ parallel to the imaginary axis with a real part obeying $|-\frac{E}{2}+\omega_k,\frac{E}{2}-\omega_k|  < \mathrm{Re}(k_0) < 3m-\frac{E}{2}$.  We will evaluate the contribution of the contour $C_1$ and show that for $E<4m$ the contribution from the contour $C_2$ to Eq.~\eqref{eq:DWBA} vanishes exponentially as $R$ increases.  The contribution from $C_1$ can be evaluated from Cauchy's theorem so that Eq.~\eqref{eq:psi-in_2} becomes
\begin{eqnarray}
\left[\KCTC\psi^\mathrm{in}_{lm}\right](k',P) &=& \frac{i}{\sqrt{4\pi}} \int\!\!\int d^3 z \frac{d^3 k}{(2\pi)^3} 
\label{eq:psi-in_3} \\
&& \hskip 0.3 in \cdot \Biggl\{\Gamma_0\Bigl(k'+\frac{P}{2},k+\frac{P}{2}\Bigr) \VCTC\bigl(|\vec z|\bigr)
                        \Gamma_0\Bigl(-k'+\frac{P}{2},-k+\frac{P}{2}\Bigr)   \nonumber \\
&& \hskip0.6 in 
\cdot \frac{e^{i(\vec k - \vec k')\cdot\vec z}}{4\omega_k E\bigl(\sqrt{k^2+m^2} -\frac{E}{2})}
 \int d\Omega_{\hat p} M_E\Bigl(k,p,P\Bigr) Y_{lm}(\hat p)\Biggr\}_{k_0=\omega_k-\frac{E}{2}}. \nonumber
\end{eqnarray}

For the previous case of $\psi^0_{lm}$ we were able to simplify this expression by introducing a charge density $\overline{\rho}(\vec r)$ and expressing each of the vertex functions $\Gamma_0(\pm k' +\frac{P}{2},\pm k +\frac{P}{2})$ as a Fourier transform of $\overline{\rho}(\vec r)$.   We can introduce the same simplification in this case as well but first we must show that for large $R$ both of the vertex functions in Eq.~\eqref{eq:psi-in_3} can be replaced by their on-shell values. 
This simplification is not immediately possible here because the integration over $k_0$ carried out when deriving Eq.~\eqref{eq:psi-in_3} evaluated only the vertex function $\Gamma_0(k' +\frac{P}{2},k +\frac{P}{2})$ on-shell.  For the other vertex function the time component of the incoming four-momentum is $-k_0+\frac{E}{2} = E-\omega_k$.  Thus, some added steps are needed to show that both vertex functions can be replaced by their on-shell values.

As a result, we will continue to work with Eq.~\eqref{eq:psi-in_3} and use Cauchy's theorem a second time to evaluate the integral over the magnitude of $\vec k$ exploiting the limit that R (which provides a lower bound on $|\vec z|$) is large.  We begin by using polar coordinates to represent the vector $\vec k$: $\vec k = k \hat k$, expressing $\vec k$ as the product of the magnitude $k$ and the unit vector $\hat k$ which is determined by the polar coordinates $\theta$ and $\phi$ in the usual way $\hat k = (\sin \theta \cos \phi, \sin \theta \sin \phi, \cos \theta)$.

Defined in this way the variable $k$ enters the amplitudes $\Gamma_0$ and $M_E$ in the exponent of a factor that performs an exponentially convergent Fourier transform. Thus, we expect these amplitudes to be analytic functions of $k$ in the region around the real line.  We will express the function $e^{i\vec k\cdot \vec z}$ in the usual series~\cite{Jackson:1998nia}:
\begin{equation}
e^{i\vec k\cdot \vec z} = \sum_{l',m'} 4\pi i^{l'} j_{l'}(kz)Y_{l'm'}(\hat z) Y_{l'm'}^*(\hat k).
\label{eq:PlaneWaveExpansion}
\end{equation}
Here $k=|\vec k|$ and $z=|\vec z|$ where the context should prevent confusion with the four vectors $k$ and $z$ .
Since $R$ is large we will replace the spherical Bessel function $j_{l'}(kz)$ by its asymptotic form 
\begin{equation}
j_{l'}(kz) \asymp \frac{e^{i(kz -\frac{\pi l'}{2})} - e^{-i(kz -\frac{\pi l'}{2})}}{2ikz}.
\end{equation}
Next we recognize that the integral in the right hand term of this equation over the interval $0 \le k < \infty$ is exactly the same as the integral of the left-hand term over the interval $-\infty > k \le 0$ allowing us to rewrite integral over $\vec k$ in Eq.~\eqref{eq:psi-in_3} in polar coordinate but with the integral over $k$ extending over the entire real line:
\begin{eqnarray}
\int d^3 k \frac{e^{i(kr -\frac{\pi l}{2})} - e^{-i(kr -\frac{\pi l}{2})}}{2ikr} 
\to \int d\Omega_{\hat k} \int_{-\infty}^\infty k^2 dk  \frac{e^{i(kr -\frac{\pi l}{2})}}{2ikr}
\end{eqnarray}
This transformation of the integral of the second term on the left-hand side of this equation to an integral over negative values of $k$ can be justified by performing a simultaneous change in the sign of the variable $k$ and the sign of the direction of $\hat k$ by performing a parity transformation on the integration variable $\hat k$.  This has no effect on the variable $\vec k = k \hat k$ appearing in Eq.~\eqref{eq:psi-in_3}.  However, in the product $j_{l'}(kz) Y_{l'm'}^*(\hat k)$ it changes the $e^{-(ikz-\frac{\pi l'}{2})} Y_{l'm'}^*(\hat k)$ term into $e^{+(ikz-\frac{\pi l'}{2})} Y_{l'm'}^*(\hat k)$.\footnote{This argument is most easily presented using the asymptotic form of $j_{l'}$.  However, it is also true if used to transform all of the terms in $j_{l'}$ with one sign of the exponent into the terms with the other sign, {\it i.e.} a transformation between the two terms that appear when $j_{l'}$ is written in terms of spherical Hankel functions of the first and second kind.}

With this change of coordinates the contribution to Eq.~\eqref{eq:psi-in_3} from the term with specific values of $l'$ and $m'$ becomes:
\begin{eqnarray}
\left[\KCTC\psi^\mathrm{in}_{lm}\right]_{l'm'}(k',P) &=& \sqrt{4\pi} i \int d^3 z \int_{-\infty}^\infty \frac{k dk d\Omega_{\hat k}}{(2\pi)^3} 4\pi e^{-i\vec k\,'\cdot z}Y_{l'm'}(\hat z)Y_{l'm'}(\hat k)^*
\label{eq:psi-in_4} \\
&& \hskip -0.7 in \cdot \Biggl\{\Gamma_0\Bigl(k'+\frac{P}{2},k+\frac{P}{2}\Bigr) \VCTC\bigl(|\vec z|\bigr)
                        \Gamma_0\Bigl(-k'+\frac{P}{2},-k+\frac{P}{2}\Bigr)   \nonumber \\
&& \hskip -0.4 in 
\cdot \frac{e^{ik |z|}}{2i|\vec z|} \frac{1}{4\omega_k E\bigl(\sqrt{k^2+m^2} -\frac{E}{2})\bigr)}
 \int d\Omega_{\hat p} M_E\Bigl(k,p,P\Bigr) Y_{lm}(\hat p)\Biggr\}_{k_0=\omega_k-\frac{E}{2}}. \nonumber
\end{eqnarray}
The integrand in Eq.~\eqref{eq:psi-in_4} contains poles at $k = \pm\sqrt{\left(\frac{E}{2}\right)^2-m^2} \pm i\epsilon$.  Because of the positive imaginary coefficient of the variable $k$ in the exponent of the integrand, we can shift the integration contour, adding a positive imaginary constant.  We can then write the expression in Eq.~\eqref{eq:psi-in_4} as the integral over this shifted horizontal contour with a positive imaginary part added to the Cauchy-theorem contribution from the pole at $k= \sqrt{\left(\frac{E}{2}\right)^2-m^2} + i\epsilon$ that must be crossed as the integration contour is shifted.

Neglecting the contribution from the shifted contour which falls exponentially with increasing $R$ (and hence increasing $|\vec z|$), we retain only the contribution from this pole:
\begin{eqnarray}
\left[\KCTC\psi^\mathrm{in}_{lm}\right]_{l'm'}(k',P) &=& \sqrt{4\pi}\frac{i}{(2\pi)^2 16\omega_p} \int d^3 z \int d\Omega_{\hat k} e^{-i\vec k\,'\cdot z}Y_{l'm'}(\hat z)Y_{l'm'}(\hat k)^*
\label{eq:psi-in_5} \\
&& \hskip 0.3 in \cdot \Biggl\{\Gamma_0\Bigl(k'+\frac{P}{2},k+\frac{P}{2}\Bigr)  \VCTC\bigl(|\vec z|\bigr)
                        \Gamma_0\Bigl(-k'+\frac{P}{2},-k+\frac{P}{2}\Bigr)   \nonumber \\
&& \hskip0.6 in 
\cdot \frac{e^{ik |\vec z|}}{|\vec z|} 
 \int d\Omega_{\hat p} M_E\Bigl(k,p,P\Bigr) Y_{lm}(\hat p)\Biggr\}_{\substack{k=\sqrt{\left(\frac{E}{2}\right)^2-m^2} \\ k_0=0\hfill}}. \nonumber
\end{eqnarray}
We should also observe that the contribution to the $k_0$ integral in Eq.~\eqref{eq:psi-in_2} coming from the displaced contour $C_2$ which we have neglected has no singularity for real values of k if $E<4m$ and hence can contribute only to a term falling exponentially in $R$ for large $R$, justifying its omission. 

We have now established the result described in Section~\ref{sec:psi_0} following Eq.~\eqref{eq:FormFactor-PS}: since the large $R$ behavior comes from the on-shell form factors, we can neglect any off-shell contribution and replace $\Gamma_0(\pm k'+\frac{P}{2},\pm k+\frac{P}{2})$ by the Fourier transform of a charge density as given in Eq.~\eqref{eq:FormFactor-PS}.

Once justified, this simplification is best made earlier in our derivation of the large $R$ behavior of the $\left[\KCTC\psi^\mathrm{in}_{lm}\right](k',P)$.  Thus, we return to Eq.~\eqref{eq:psi-in_3} but now written in terms of the charge density:
\begin{eqnarray}
\left[\KCTC\psi^\mathrm{in}_{lm}\right](k',P) &=& \frac{iE^2}{\sqrt{4\pi}}   \int\!\!\int\!\!\int\!\!\int d^3 z\, d^3 r_2\, d^3 r_1 \frac{d^3 k}{(2\pi)^3} 
\label{eq:psi-in_6} \\
&& \hskip 0.3 in \cdot \Biggl\{e^{i(\vec k - \vec k\,')\cdot \vec r_2}\overline{\rho}(\vec r_2) \VCTC\bigl(|\vec z|\bigr)
                       e^{-i(\vec k - \vec k\,')\cdot \vec r_1}\overline{\rho}(\vec r_1)   \nonumber \\
&& \hskip0.6 in 
\cdot \frac{e^{i(\vec k - \vec k')\cdot\vec z}}{2\omega_k E\bigl(\sqrt{k^2+m^2} -\frac{E}{2})}
 \int d\Omega_{\hat p} M_E\Bigl(k,p,P\Bigr) Y_{lm}(\hat p)\Biggr\}_{k_0=\omega_k-\frac{E}{2}}. \nonumber
\end{eqnarray}
We then repeat the steps used to set $|\vec k| = |\vec p|$ but will work with the combined plane wave factor that now appears in Eq.~\eqref{eq:psi-in_6}:
\begin{equation}
e^{i(\vec k - \vec k\,')\cdot(\vec r_2+\vec z -\vec r_1)} \equiv e^{i(\vec k - \vec k\,')\cdot \vec w},
\end{equation}
expanding $e^{i\vec k\cdot \vec w}$ in spherical harmonics, combining the terms in $j_l(kw)$ so that the $k$ contour runs from $-\infty$ to $+\infty$, shifting the $k$ contour into the positive imaginary half-plane and keeping the $k=+|\vec p|$ pole. 

With this altered treatment the vector $\vec k$ appears only in the factor $e^{i\vec k\cdot \vec w}$ and the scattering amplitude $M_E(k,p,P)$.  When we expand the factor $e^{i\vec k\cdot \vec w}$ in spherical harmonics the angular integral over the direction of $\vec k$ then selects only the term where what was labeled $l'$ in Eq.~\eqref{eq:psi-in_5} now must equal $l$, assuming that the scattering amplitude $M_E(k,p,P)$ is rotationally symmetric.  The resulting simpler expression becomes:
\begin{eqnarray}
\left[\KCTC\psi^\mathrm{in}_{lm}\right](k',P) &=& \frac{iE}{4\sqrt{\pi}}   
\int d^3w\, e^{-i\vec k\,'\cdot\vec w} Y_{lm}(\hat w)
\label{eq:psi-in-result} \\
&& \hskip 0.3 in \cdot\, \frac{e^{ipw}}{w} \left\{\int\!\!\int d^3 r_2\, d^3 r_1 \overline{\rho}(r_2) \VCTC(\vec w - \vec r_2 + \vec r_1) \overline{\rho}(r_1) \right\}  M_l(E). \nonumber
\end{eqnarray}

We can now combine  these results for $\left[\KCTC\psi^{(0)}_{lm}\right]$ and $\left[\KCTC\psi^\mathrm{in}_{lm}\right]_{l'm'}$ by adding the results given in Eqs.~\eqref{eq:psi0-result} and \eqref{eq:psi-in-result}:
\begin{eqnarray}
\left[\KCTC\Psi^\mathrm{in}_{lm}\right]_{l'm'}(k',P) &=& -\int d^3 w  e^{-i\vec k\,'\cdot w}Y_{lm}(\hat w)
\label{eq:psi-tot_1} \\
&& \hskip 0.3 in \cdot \left\{\int\!\!\int d^3 r_2\, d^3 r_1 \overline{\rho}(r_2) \VCTC(\vec w - \vec r_2 + \vec r_1) \overline{\rho}(r_1) \right\}  \nonumber \\
&& \hskip0.6 in 
\cdot \sqrt{4\pi}E^2 \left\{2\frac{i^l\sin(p|\vec w|-\frac{\pi l}{2})}{p|\vec w|}+\frac{1}{16\pi\omega_p}\frac{e^{ip |\vec w|}}{|\vec w|}\left[32\pi\frac{\omega_p}{p}\frac{e^{2i\delta_l}-1}{2i}\right] \right\}. \nonumber \\
&& \hskip -1.3 in = -4\sqrt{\pi}E^2\int d^3 w  e^{-i\vec k\,'\cdot w} Y_{lm}(\hat w)
\label{eq:psi-tot_2} \\
&& \hskip -1.0 in \cdot  \left\{\int\!\!\int d^3 r_2\, d^3 r_1 \overline{\rho}(r_2) \VCTC(\vec w - \vec r_2 + \vec r_1) \overline{\rho}(r_1) \right\}
\frac{e^{i\delta_l} i^l \sin\bigl(p|\vec w|-\frac{\pi l}{2} + \delta_l\bigr)}{p|\vec w|}. \nonumber
\end{eqnarray}

The full result including the both the factors $\Psi^\mathrm{out}(k',P)$ on the left and $\Psi^\mathrm{in}(k,P)$ on the right as in Eq.~\eqref{eq:DWBA} and choosing the first-order term in an expansion of Eq.~\eqref{eq:phase-shift} in powers of $\alpha$ is given by:
\begin{eqnarray}
\dCTC_l &=& \frac{p}{32\pi\omega_p} {\MCTC}_{,l} e^{-2\delta_l} \\
    &=& -p\omega_p\int_0^\infty w^2 dw 
\left\{\int\!\!\int d^3 r_2\, d^3 r_1 \overline{\rho}(r_2) \VCTC(\vec w - \vec r_2 + \vec r_1) \overline{\rho}(r_1) \right\} \nonumber \\
&& \hskip 0.5 in \cdot \Bigl[\cos{\delta_l}j_l(pw) + \sin(\delta_l)n_l(pw)\Bigr]^2 e^{-\mu w}  \label{eq:result}
\end{eqnarray}
Here we have replaced the leading asymptotic behavior of the spherical Bessel functions $j_l(pw)$ and $n_l(pw)$ by those functions without approximation to reinstate the complete result of our derivation, restoring terms which fall as powers of $1/R$ but which had been omitted for clarity when $l>0$.

We have also introduced the factor $\exp\{-\mu w\}$ which depends on the screening mass $\mu$ into Eq.~\ref{eq:result} to regulate the long-distance logarithmic singularity that would otherwise be present in the integral over $w$.  This singularity is the manifestation in perturbation theory of the long-range character of the Coulomb potential which appears in exact solutions of the Schrodinger and Dirac equations as a common phase which depends logarithmically on the radial variable $w$.  Since this is a common phase, equal for all partial waves, it does not enter most physical quantities which implies that for small $\mu$, the $\mu$-dependence introduced arbitrarily into Eq.~\ref{eq:result} will also not appear in the final physical quantities that we wish to calculate.  For example, the quantity $\eta_{+-}$ that measures CP violation in the neutral kaon decay into two charged pions~\cite{Donoghue:1992dd} is a ratio of decay amplitudes into a common $\pi^+\pi^-$ final state from which this phase will naturally cancel.

Equation~\eqref{eq:result} provides the analytic method to calculate the contribution of $\VCTC$ to the $\pi^+\pi^+$ scattering phase shift in terms of physical QCD properties: the $\pi\pi$ scattering phase shift $\delta_l$ and the Fourier transform $\overline{\rho}(|\vec r|)$ of the $\pi^+$ electromagnetic form factor $F(q^2)$.  The result applies to the fully relativistic case provided the center-of-mass energy $E<4m$ and that $R$ is sufficiently large that omitted terms falling exponentially in $R$ can be omitted.  The $\pi\pi$ scattering phase shift $\delta_l$ can be determined by the usual application of L\"uscher's finite-volume quantization methods.  The convolution of two factors of the charge density with the potential $\VCTC$ can be most easily determined by expanding the function $\VCTC(\vec w +\vec r_2 -\vec r_1)$ assuming $|\vec w| \gg |\vec r_i|$.  The leading term is determined by the pion's charge and the $1/R^2$ terms by its charge radius.

\section{Conclusion}
\label{sec:conclusion}

As a first step in calculating the electromagnetic and $m_u-m_d$ contributions to the measure of direct CP violation $\varepsilon'$ we have presented in detail a method to determine the contribution of the Coulomb potential to the $\pi^+\pi^+$ scattering phase shift.  If the quantization of QED is carried out in Coulomb gauge with $\vec \nabla \cdot \vec A = 0$, this is well-defined and will give the complete E\&M contribution when the effects of transverse radiation are included. 

The calculation of these Coulomb effects is itself separated into two parts: one in which the separation $r$ of the two charge operators in the Coulomb energy is less than $R$ (referred to as the truncated Coulomb potential, $\VTC$)
and a second in which this separation is greater than $R$, $\VCTC$.  The effects of $\VTC$ can be directly determined from a finite-volume, lattice QCD calculation while those of $\VCTC$ can be obtained from an analytic expression which we derive.   

Given the finite range of QCD and the truncated potential $\VTC$ their contributions to the $\pi^+\pi^+$ scattering phase shift can be computed with the standard finite-volume methods giving a result with finite-volume errors which fall exponential in the spatial extent $L$ of the volume, with the exception of power-law corrections asring from the usual neglect of scattering phase shifts for angular momenta $l \ge 4$.  Similarly, the analytic expression given in Eq.~\eqref{eq:result} determines the contribution of $\VCTC$ to exponential accuracy in the truncation radius $R$ in terms of the $\pi^+\pi^+$ scattering phase shift without E\&M corrections and the pion form factor.  This control of the power-law finite-volume corrections may be important when exploiting finite-volume quantization in which the results themselves are power-law, finite-volume effects.

In two appendices we discuss the use of the \qedl approach to E\&M corrections again considering the Coulomb potential but in the non-relativistic limit.  We present a derivation of a \qedl, finite-volume quantization condition which determines the E\&M corrections to scattering phase shift that does not involve an effective range approximation and provides an alternative to the earlier treatment of Bean and Savage~\cite{Beane:2014qha}.  We also carry out a numerical study which suggests that this \qedl approach gives accurate results in spite of the presence of uncontrolled $1/L^3$ corrections.

Two important future steps are required for a complete lattice calculation of the electromagnetic and $m_u-m_d$ contributions to $\varepsilon'$.  First the current Coulomb potential treatment must be generalized to the two-channel $I=0$ and $I=2$ $\pi\pi$ system and the $K\to\pi\pi$ decay analyzed.  This was outlined in Ref.~\cite{Christ:2017pze} for the non-relativistic case.  Second the contribution of transverse radiation must be included which requires the usual treatment of infrared divergences and the effects of three-particle $\pi\pi\gamma$ states.  Each is the subject of on-going study.

\section*{Acknowledgments}

We would like to thank our RBC and UKQCD collaborators for their ideas and assistance.  The work of N.H.C., J.M.K. and T.H.N. was supported in part by U.S. DOE grant \#DE-SC0011941 while X.F. was supported in part by NSFC of China under Grant No. 11775002 and No. 12070131001 and National Key Research and Development Program of China under Contract No. 2020YFA0406400.

\appendix
\section{\qedl finite-volume quantization}
\label{sec:QEDL-quantization}

The \qedl finite-volume formulation of QED~\cite{Hayakawa:2008an} is widely used in lattice QCD calculations to include the effects of E\&M.  In this approach the Coulomb potential or photon propagator is written as a Fourier series appropriate for a function defined in a finite volume and obeying periodic boundary conditions.  The singular modes at zero wave number are discarded.  The resulting Coulomb potential or photon propagator is convenient to use in such a volume but differs from the corresponding physical quantity by terms which behave as $1/L^n$ for $n \ge 1$.  These power law imperfections in the \qedl formulation lead to lattice results with finite-volume errors that also fall as powers of $1/L$ with the first few terms taking universal, point-charge values.  However, those errors falling with higher powers of $1/L$ will be unknown and must be removed by studying multiple volumes and extrapolating $L \to\infty$.  For a recent reference on this topic see the paper of Davoudi, {\it et al.}~\cite{Davoudi:2018qpl}.

For a problem of the sort being studied here in which the quantization of finite-volume energies plays an essential role, the need when using \qedl to extrapolate $L \to\infty$ creates potentially serious difficulties.  This motivates the truncated Coulomb potential approach developed in this paper in which such new power-law finite-volume corrections are avoided.  Never-the-less, given the frequent use of $\QEDL$, we present in this appendix a non-relativistic derivation of a finite-volume quantization condition that might be used in a \qedl calculation to determine the E\&M corrections to the low-energy $\pi^+\pi^+$ scattering phase shift.  In the following Appendix~\ref{sec:QEDL-numerics}, we investigate its accuracy in a numerical study of a simple potential model.  For the example studied, we find accurate results for finite $L$ without the need for an $L\to\infty$ extrapolation.  

The derivation given in this appendix can be viewed as an alternative to that presented by Beane and Savage~\cite{Beane:2014qha} and used by Beane, {\it et al.}~\cite{NPLQCD:2020ozd}.  In contrast to Ref.~\cite{Beane:2014qha}, we express the quantization condition directly as an explicit formula for the resulting $\pi^+\pi^+$ phase shift rather than as a condition on the scattering length and effective range, defined in a fashion appropriate for a Coulomb scattering problem, which would correspond to that phase shift.

\subsection{Properties of the \qedl approximation to the Coulomb potential}

The \qedl potential is simply the Fourier-transformed Coulomb potential in a finite volume with the zero-mode omitted:
\beq\label{eq:qedlcoul}
V_{{\rm QED}_{\rm L}}({\vec r}) = \frac{e^2}{L^3}\sum_{{\vec n} \neq {\vec 0}} \frac{e^{i {\vec p_{\vec n}}\cdot {\vec r}}}{p_{\vec n}^2} \,,
\eeq
where $\vec p_{\vec n} = \frac{2\pi}{L}(n_1,n_2,n_3)$, with the integers $n_i$, $1 \le i \le 3$ giving the three components of the vector $\vec n$ and $\vec 0 = (0,0,0)$ is the zero mode.

We are interested in the $l = 0$ component of this potential near the origin and how it deviates from the Coulomb potential, $e^2/4\pi r$, at short distances. This is most easily done by noting the similarities between the \qedl potential and the periodic Helmholtz function $G_0$, introduced in L\"uscher's discussion~\cite{Luscher:1990ux} of finite-volume energy quantization: 
\beq\label{eq:qedlcoul}
V_{{\rm QED}_{\rm L}}({\vec r}) = \lim_{k \to 0} - e^2 \left[G_0({\vec r}; k) - \frac{1}{L^3k^2}\right] \,,
\eeq
where
\beq\label{eq:helmholtzgreens}
G_0({\vec r};k) = \frac{1}{L^3}\sum_{\vec n} \frac{e^{i {\vec p_{\vec n}}\cdot {\vec r}}}{k^2 - p_{\vec n}^2}\,.
\eeq

Substituting the known spherical harmonic expansion of the periodic Helmholtz function and evaluating the limit of vanishing $k$, we obtain the standard result:
\beq\label{eq:qedl-small-r}
V_{{\rm QED}_{\rm L}}({\vec r}) = \frac{e^2}{4\pi} \left[ \frac{1}{r} + \frac{Z_{00}(1;0)}{\pi^{3/2}(L/2\pi)} + \frac{2\pi r^2}{3L^3}  + \mathcal{O}\left(\frac{r^4}{L^5}Y_{l=4}({\hat r})\right) \right]\,,
\eeq
where $Z_{00}$ is the standard zeta function:
\begin{equation}
Z_{00}(s,q^2) = \frac{1}{\sqrt{4\pi}}\sum_{\vec n}\frac{1}{(n^2-q^2)^s}.
\label{eq:Z00}
\end{equation}
The difference between the Coulomb potential and the \qedl potential is plotted in Fig.~\ref{fig:qedl} for $L=4$. Even when approaching the boundary of the box, about which the \qedl potential is symmetric, the difference between the \qedl and Coulomb potentials is well represented by the two leading correction terms in Eq.~\eqref{eq:qedl-small-r} with the largest deviations seen in those directions where the greatest distance from the origin can be reached.  However, small direction-dependent discrepancies can also be seen.

{\embr
\begin{figure}[!htp]
\centering
\includegraphics[width=0.45\textwidth]{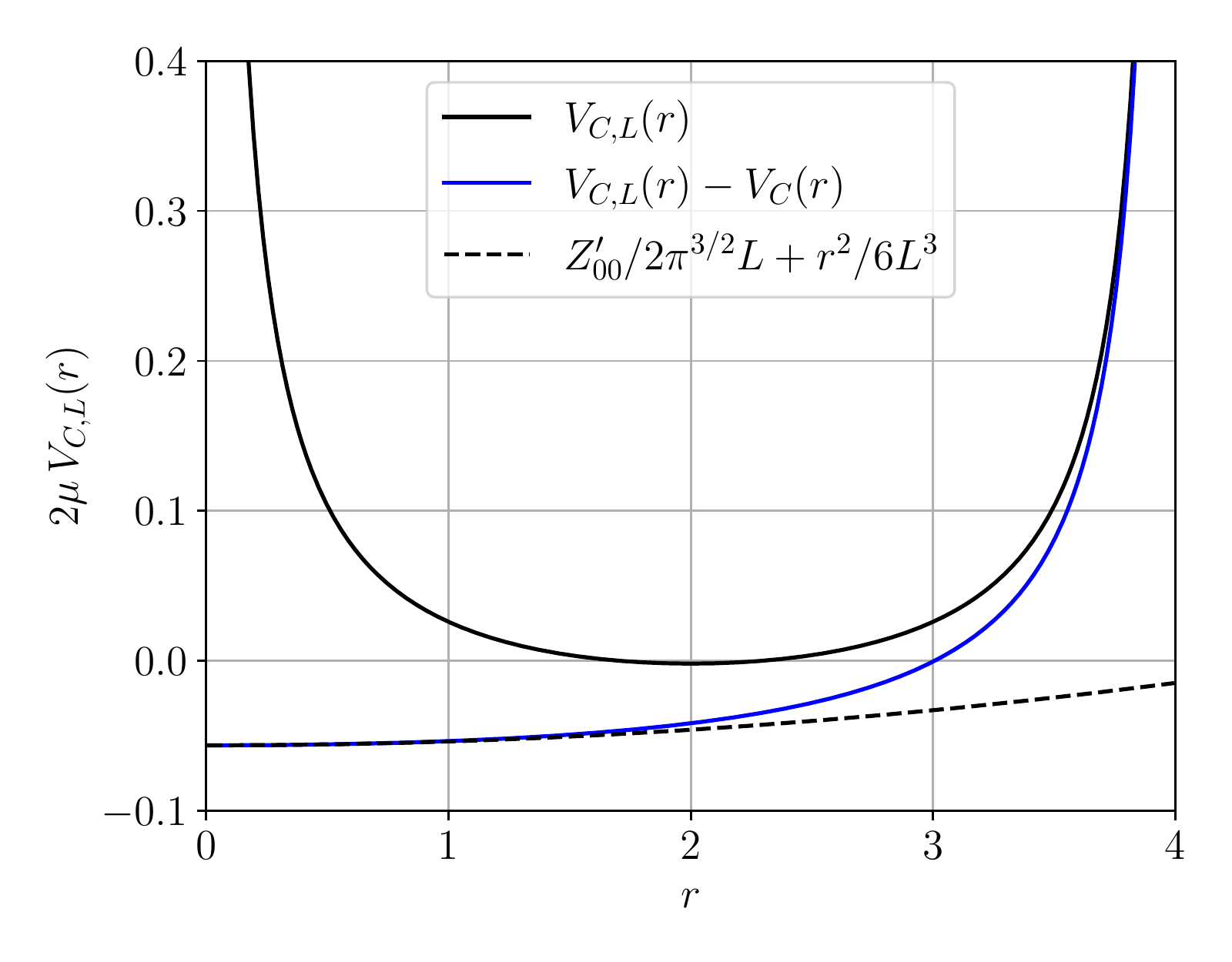}
\includegraphics[width=0.45\textwidth]{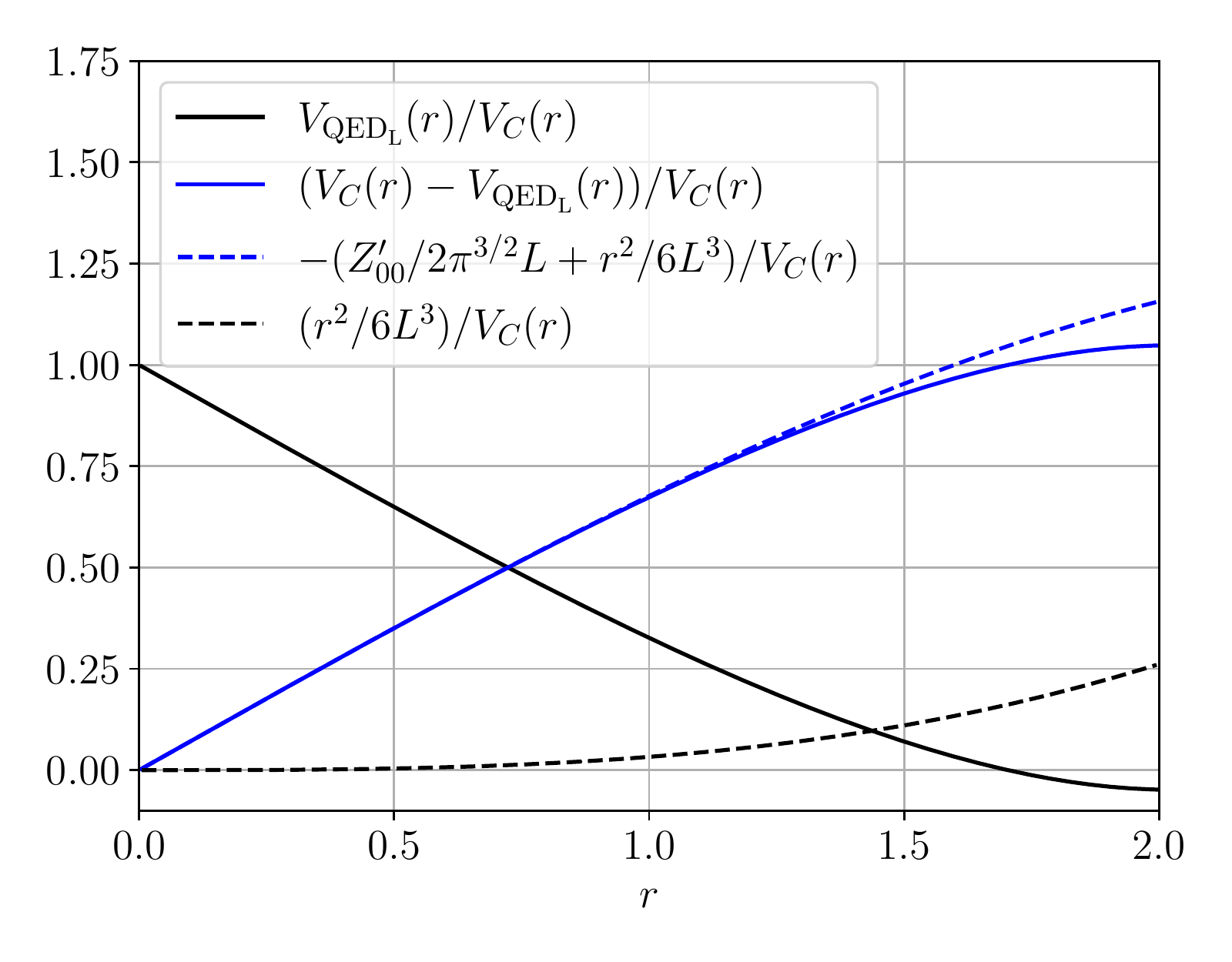}
\includegraphics[width=0.45\textwidth]{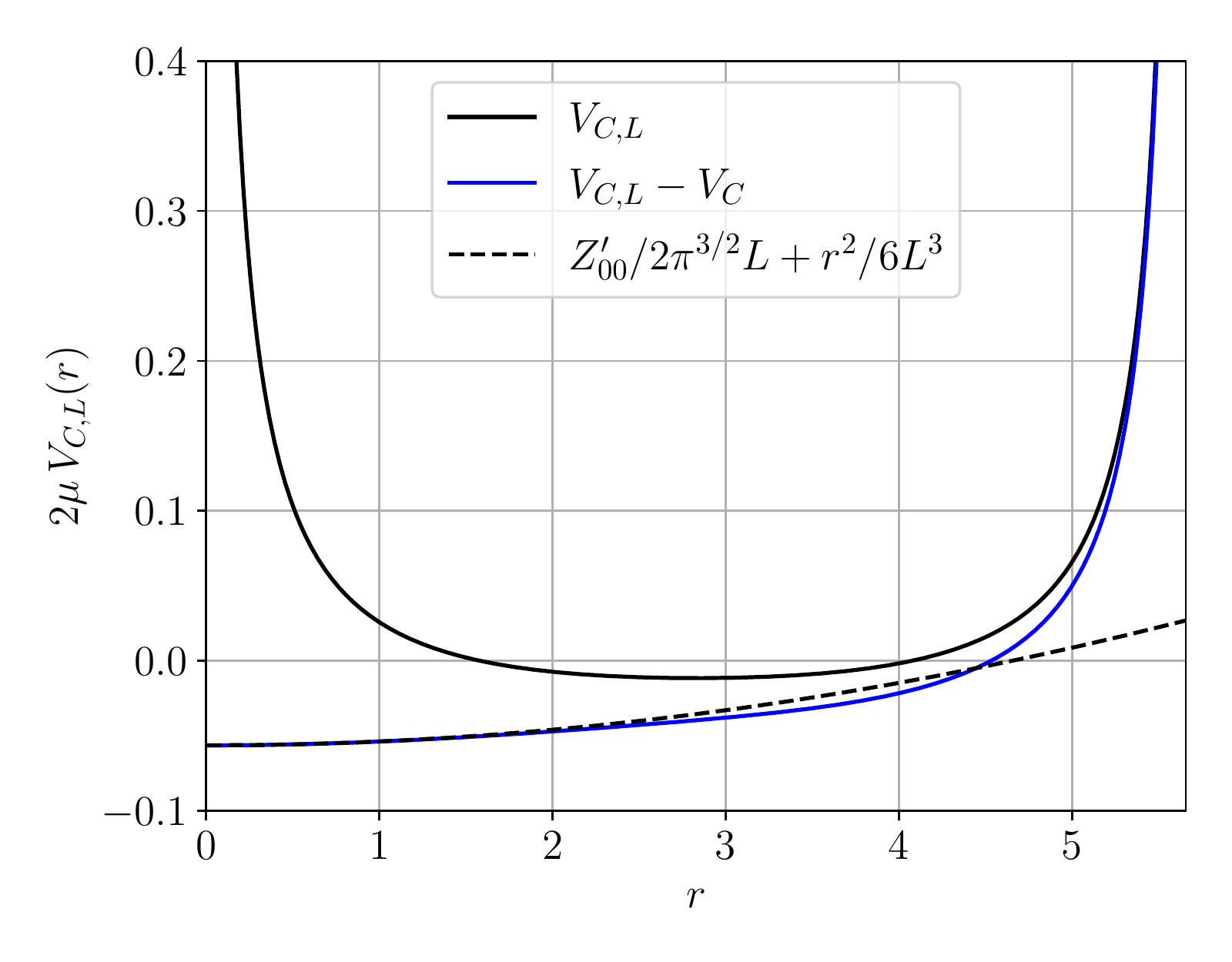}
\includegraphics[width=0.45\textwidth]{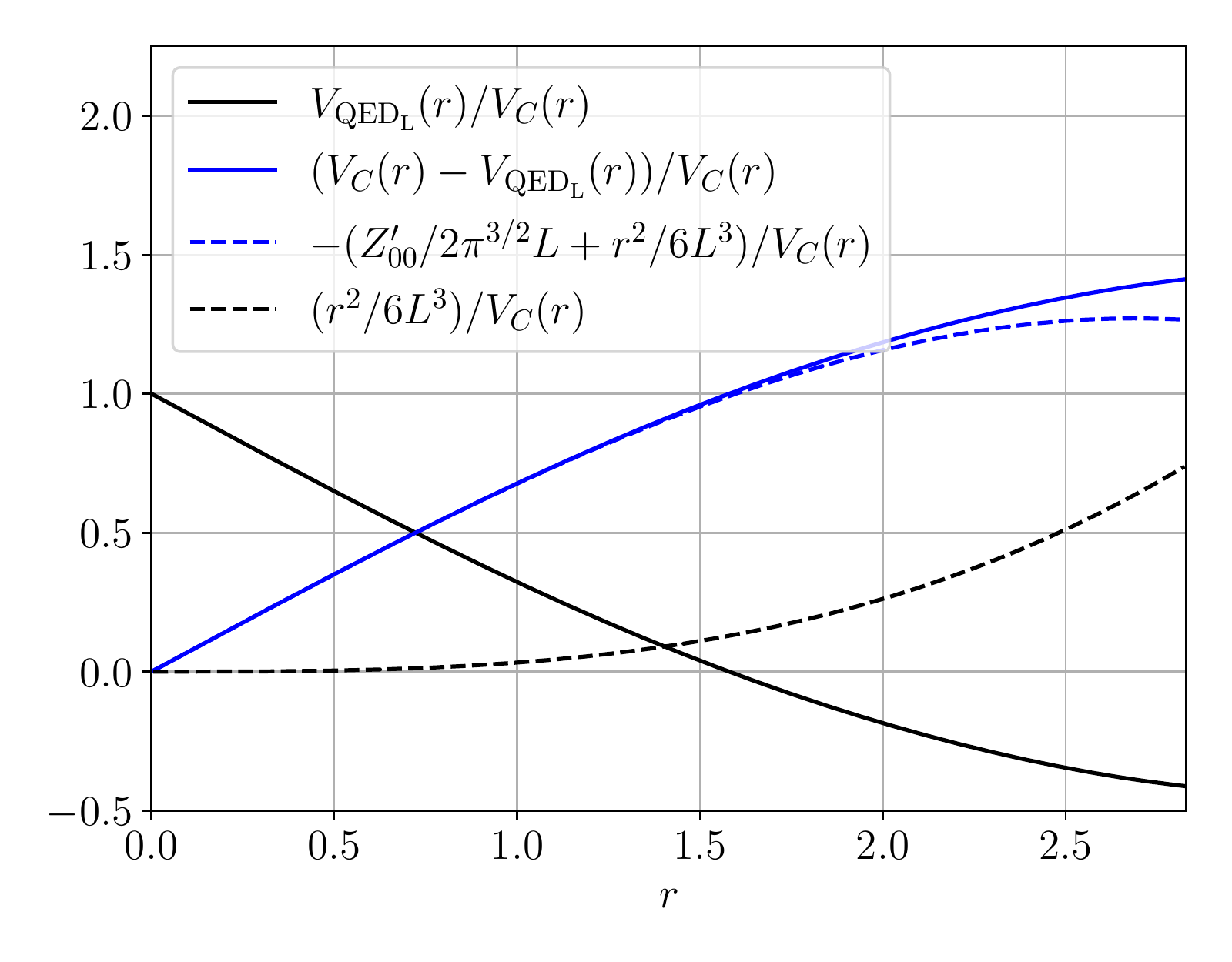}
\includegraphics[width=0.45\textwidth]{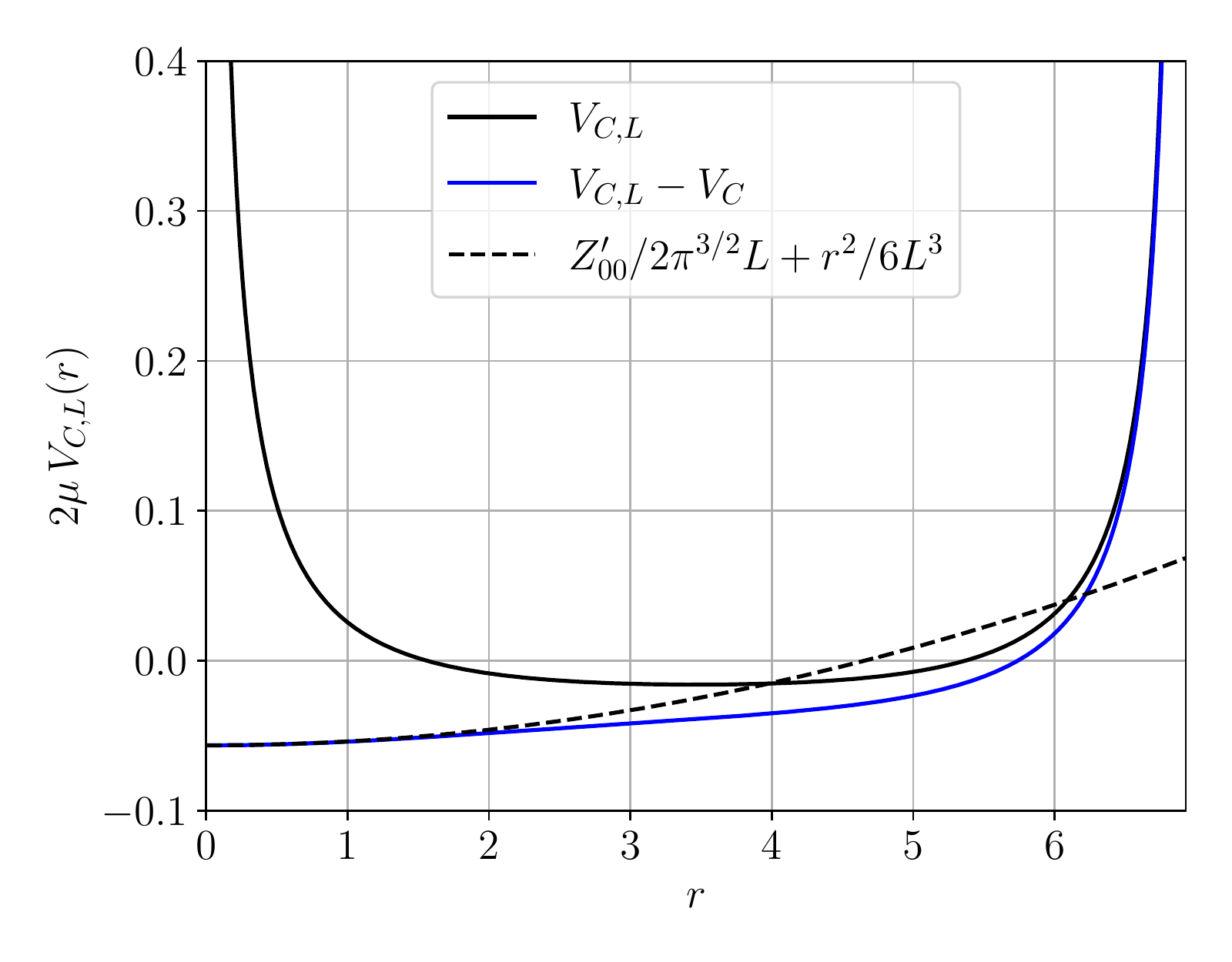}
\includegraphics[width=0.45\textwidth]{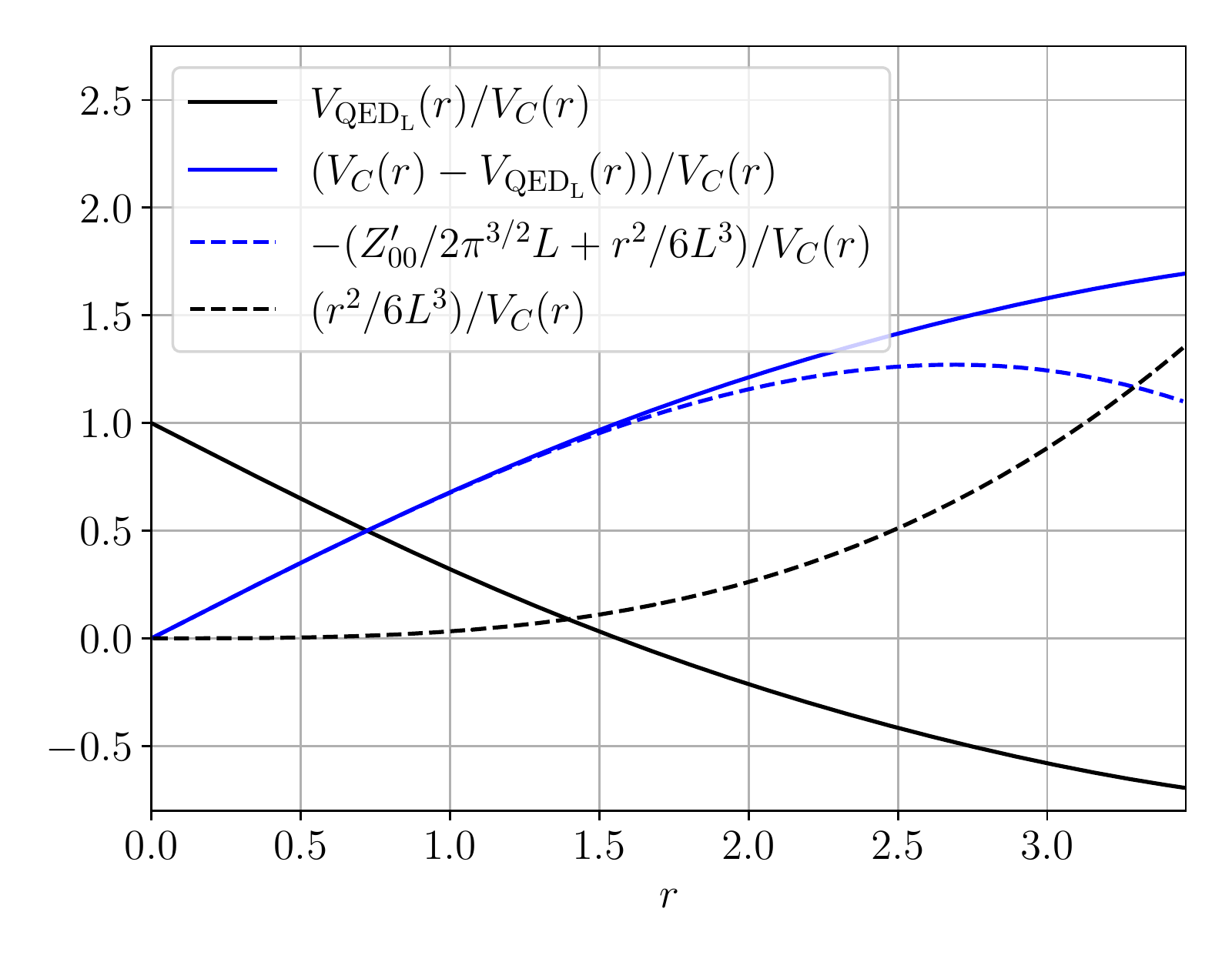}
\caption{\label{fig:qedl} Left: The \qedl potential (solid black line), its difference from the Coulomb potential (solid blue line), and the two leading correction terms of Eq.~\eqref{eq:qedl-small-r} (dashed black line) in a box of size $L=4$.  Right: The \qedl potential (solid black line), its difference from the Coulomb potential (solid blue line), the two leading correction terms in Eq.~\eqref{eq:qedl-small-r} (dashed blue line), and the $r^2$ correction term alone (dashed black line), each divided the Coulomb potential $V_C$.  For the upper, middle and lower rows $r$ is the distance from the origin along a line parallel to an edge, a face diagonal or a body diagonal, respectively.}
\end{figure}
} 

From Eq.~\eqref{eq:qedl-small-r} we see that near the origin the \qedl potential resembles the classical $e^2/4\pi r$ Coulomb potential plus additional power-law terms. The second term acts as an energy shift whose effects in the problem at hand can be completely removed.  This term corresponds to the universal $1/L$ term in QED self-energy calculations~\cite{Borsanyi:2014jba, Davoudi:2018qpl}.  The third term can be recognized as the potential energy due to a uniform charge distribution making the total charge in the volume zero.  This and higher-order unphysical terms will combine with the finite-range strong interactions to introduce ''structure-dependent" errors which can be removed by an explicit $L\to\infty$ extrapolation.  These are related to the structure-dependent $1/L^3$ terms found in \qedl self-energy calculations~\cite{Borsanyi:2014jba, Davoudi:2018qpl}.

\subsection{L\"uscher finite-volume quantization extended to include \qedl}

Following this brief description of the properties of the $\VQEDL(r)$ we will use this potential and derive an approximate quantization condition which relates the energy eigenvalues of two-particle eigenstates of the combined QCD and \qedl Hamiltonian in a periodic volume of side $L$ to the infinite-volume scattering phase shift that results from their combined effects.  In contrast with the earlier sections of this paper, we consider the scattering of two scalar particles in the non-relativistic limit.  In this case, our use of the Coulomb potential and neglect of transverse radiation can be justified as giving the leading order contribution in a non-relativistic expansion.  

We follow the standard approach of treating an alternative non-relativistic problem in which the QCD interaction is replaced by a simple scattering potential $V_S(r)$ in non-relativistic quantum mechanics.  Since the relation obtained between finite-volume energies and the scattering phase shifts arises from the constraint that the large $r$ behavior of the scattering solution obeys periodic boundary conditions in a finite volume, we expect this same relation to hold for QCD where the large $r$ behavior of the scattering states depends on the scattering phase shifts in the same way.

The result described here is a generalization of that obtained by L\"uscher to include the effects of the Coulomb potential and the approach used in its derivation follows closely that of Ref.~\cite{Luscher:1990ux}.  This problem was studied earlier by Beane and Savage~\cite{Beane:2014qha} who present a similar quantization condition specialized to the low energy region where an effective range approximation is valid.  However, the method used by Beane and Savage is general and could be used to obtain the result presented here.  

In fact, the derivation given here and that of Bean and Savage differ in the same way as do the treatment presented by L\"uscher in Ref.~\cite{Luscher:1990ux} and that of van Baal~\cite{vanBaal:1990mp}. While L\"uscher's treatment matches the asymptotic form of the full "strong-interaction" solution to the behavior of the finite-volume Helmholtz equation solution outside the range of the strong-interaction potential, van Ball, following Huang and Yang~\cite{Huang:1957im}, replaces that potential by a simpler point-like pseudo-potential creating a problem which can be solved explicitly in a finite, cubic volume. The resulting relationship between the quantized finite-volume energy in this psuedo-potential problem and the phase shift predicted by that pseudo-potential is precisely the usual one.

Adding the Coulomb potential significantly changes the quantized finite-volume energies because of the long range behavior of the Coulomb potential.  Further, as discussed above, the Coulomb potential itself must be modified if it is to be consistent with the introduction of periodic boundary conditions.  For simplicity, we will consider the case where only the $s$-wave phase shift that appears in the infinite-volume scattering problem defined by the potential $V_S(r)$ alone (without the Coulomb potential) is non-zero.

Specifically, we consider the finite-volume energy eigenstates $\psi^V_E(\vec r)$ of the non-relativistic quantum mechanical Hamiltonian:
\be
\left(\frac{-\nabla^2}{2\mu} + V_S(r) + \VQEDL(r)\right)\psi^V_E(\vec{r})=E\psi^V_E(\vec r),
\label{eq:SE}
\ee
where $V_S(r)$ is a finite-range potential which plays the role of the strong interaction between two identical spin-zero particles with reduced mass $\mu$.  We examine the behavior of the wave function $\psi^V_E$ in a region where $r$ lies outside of the region of radius $r_0$ in which $V_S(r)$ is non-zero, $r > r_0$ and, to reduce the $(r/L)^n$ errors introduced by using $\QEDL$, we also require $r \ll L$ where $L$ is the length of the spatial side of the finite volume.  Following L\"uscher we consider two descriptions of the finite-volume energy eigenstate $\psi^V_E(\vec r)$ in this region.

The first is given by the Green's function $G_{\QEDL}(\vec r)$ which obeys a Helmholtz equation which now includes the \qedl potential:
\begin{equation}
\left(\nabla^2 + k^2 - 2\mu\VQEDL(r)\right) G_{\QEDL}(\vec r) = \delta(\vec r)
\label{eq:HZ} 
\end{equation}
If the function $G_{\QEDL}(\vec r)$ is expanded in spherical harmonics, those components with $l > 0$ will be regular in the entire volume $V$ and should agree with corresponding terms in a partial wave expansion of $\psi^V_E(\vec r)$ provided $E=k^2/2\mu$ since we are assuming that for $l>0$ the phase shifts $\delta_l$ induced by $V_S(r)$ are zero.  Thus, for $l>0$ we expect that the contributions to $G_{\QEDL}(\vec r)$ with angular momentum $l$ will solve Eq.~\eqref{eq:SE} throughout the volume $V$.  The $l=0$ component of $G_{\QEDL}(\vec r)$ will not solve Eq.~\eqref{eq:SE} with non-zero $V_S(r)$ and will be singular as $r\to0$.  In this appendix, we will use $G_{\QEDL}$ evaluated to first order in $\alpha$.  

The second description of the finite-volume energy eigenstate $\psi^V_E(\vec r)$ in the region $r_0\le r \ll L$ is provided by the spherical Coulomb functions $\mathcal{F}_l$ and $\mathcal{G}_l$ which solve the Schr\"odinger equation including the physical Coulomb potential and replace the spherical Bessel functions $j_l(kr)$ and $n_l(kr)$ which appear in L\"uscher's treatment.  Here we assume that for $r/L$ sufficiently small we can ignore the difference between the potential $\VQEDL$ which appears in Eqs.~\eqref{eq:SE} and \eqref{eq:HZ} and the usual Coulomb potential which appears in the equations obeyed by $\mathcal{F}_l$ and $\mathcal{G}_l$.  These functions are known to all orders in a perturbation expansion in $e^2$ and can be easily simplified to a first-order form.  

The spherical Coulomb functions appear in the $r \ge r_0$ description of the infinite-volume scattering solutions $\psi_E(r)$ and obey Eq.~\eqref{eq:SE} if $\VQEDL$ is replaced by $e^2/(4\pi r)$.  Specifically if we consider an infinite-volume energy eigenstate $\psi_{E,l}(r)$ with energy $E$ and orbital angular momentum $l$, for $r\ge r_0$ it can be written:
\be
\psi_{E,l}(r)=a_l\left\{\cos\bar{\delta}_l\,
\mathcal{F}_l(\eta,kr)+\sin\bar{\delta}_l\, \mathcal{G}_l(\eta,kr)\right\}.
\label{eq:inf-volume-ev}
\ee
where the Sommerfield parameter $\eta = \mu e^2/(4\pi k)$, $\bar{\delta}_l$ is the scattering phase shift for this combined Coulomb plus strong interaction problem and $a_l$ is a normalization factor. Recall that the large $r$ behaviors of $\mathcal{F}_l$ and $\mathcal{G}_l$ are determined by their asymptotic forms:
\begin{eqnarray}
\mathcal{F}_l(\eta,\rho) &\sim&  \sin\left(\rho-\eta\ln(2\rho) -\frac{1}{2}\pi l +\sigma_l  \right)
\label{eq:large_r0}\\
\mathcal{G}_l(\eta,\rho) &\sim& \cos\left(\rho-\eta\ln(2\rho) -\frac{1}{2}\pi l +\sigma_l  \right)
\label{eq:large_r1}
\end{eqnarray}
where
\begin{equation}
\sigma_l = \mathrm{arg}\Bigl(\Gamma(l+1+i\eta)\Bigr).
\label{eq:sigma_l}
\end{equation}
with the usual $\Gamma$ function
\begin{equation}
\Gamma(z) = \int_0^\infty dt e^{-t}t^{z-1}.
\label{eq:gamma}
\end{equation}

Thus, for large $r$
\begin{eqnarray}
\psi_{E,l}(r) &\sim& \cos{\bar{\delta}_l}\mathcal{F}_l(\eta,kr)+\sin{\bar{\delta}_l}\mathcal{G}_l(\eta,kr)
\label{eq:delta_def0} \\          
                   &  =   & \sin\left(kr-\eta\ln(2kr) -\frac{1}{2}\pi l +\sigma_l  +\bar\delta_l \right).
\label{eq:delta_def2}
\end{eqnarray}
The $\ln(2kr)$ dependence present in the large-$r$ behavior of $\psi_{E,l}(r)$ appears in the functions $\mathcal{F}_l(\eta,kr)$ and $\mathcal{G}_l(\eta,kr)$ so that $\bar{\delta}_l$ is a well-defined phase that is conventionally used to describe the remaining combined effects of the ``strong" and Coulomb  interactions.  The phase shift $\bar{\delta}_l$ is the infinite-volume quantity that we would like to determine from a finite-volume calculation.

Our Coulomb finite-volume quantization condition is then obtained by requiring that these two descriptions of the $s$-wave component of finite-volume solution $\psi^V_E(\vec r)$ agree in the region where both are valid: $r_0 \le r \ll L$.  If we neglect the differences between $\VQEDL$ and $e^2/(4\pi r)$, then these two representations of the $s$-wave part of the  finite-volume solution $\psi^V_E(\vec r)$ must be the same and the most accessible way to compare them is to extend both into the region near $r=0$ where, since they are identical, they must have the same ratio of regular to singular parts.  Thus, we must examine the small $r$ behavior of the $l=0$ component of the Green's function $G_{\QEDL}(\vec r)$ determined by Eq.~\eqref{eq:HZ} and the spherical Coulomb functions $\mathcal{F}_0$ and $\mathcal{G}_0$ in Eq.~\eqref{eq:inf-volume-ev}.  Equating the ratios of the regular and singular parts in each description will then provide the quantization condition determining the phase shift $\bar \delta_0$.   

In Eq.~\eqref{eq:qedl-small-r} we show the $1/L$ and $1/L^3$ terms that appear for small $r/L$ in the difference between the Coulomb potential and $\VQEDL$.  Both terms are potential sources of error in the phase shifts determined from this quantization condition.  While we do not attempt to characterize the errors introduced by the $r^2/L^3$ term in Eq.~\eqref{eq:qedl-small-r}, we can easily remove the $1/L$ error by equating the ratio of the singular and regular $s$-wave parts of the \qedl Helmholtz solution $G_E(\vec r)$ evaluated at the finite-volume energy $E$ with those of the spherical Coulomb functions $\mathcal{F}_0$ and $\mathcal{G}_0$ combined with the phase shift $\bar\delta$ as in Eq.~\eqref{eq:inf-volume-ev} evaluated at the energy $E'=E+\Delta E$ where
\begin{equation}
\Delta E = -\frac{e^2 Z_{00}(1;0)}{2\pi^{3/2}L}\,.
\label{eq:DeltaE}
\end{equation}
In the remainder of the appendix, we will distinguish with primes the energy and momentum ($E'$ and $k'$) that appear in the infinite-volume, Coulomb-potential quantities from the energy and momentum ($E$ and $k$) that appear in the corresponding finite-volume quantities.

The small $r$ behavior of the spherical Coulomb functions $\mathcal{F}_l$ and $\mathcal{G}_l$ is well known~\cite{NIST:DLMF} and for $l=0$, $\rho= k'r$ and $\eta'=e^2\mu/(4\pi k')$ can be written:
\begin{eqnarray}
\mathcal{F}_0(\eta',\rho)  &\sim& c_0\rho\left[1+\eta'\rho + \cdots\right] 
\label{eq:F-small_r} \\
\mathcal{G}_0(\eta',\rho) &\sim& \frac{1}{c_0}\Bigl[1+2\eta'\rho\bigl\{\ln(2\rho) + \mathrm{Re}\left[\psi(i\eta')\right] + 2\gamma_E-1 \bigr\} + \cdots\Bigr]
\label{eq:G-small_r} 
\end{eqnarray}
where
\begin{eqnarray}
c_0 &=& \left(\frac{2\pi\eta'}{e^{2\pi\eta'}-1}\right)^{\frac{1}{2}},
\end{eqnarray}
$\psi(z)=\Gamma'(z)/\Gamma(z)$ is the digamma function and $\gamma_E=0.57721\ldots$ is Euler's constant.

Next we examine the finite-volume Helmholtz equation solution $G_{\QEDL}(\vec r)$ and determine its behavior for small $r$.  As a first step we rewrite the differential equation, Eq.~\eqref{eq:HZ} obeyed by $G_{\QEDL}(\vec r)$ as the Lipman-Schwinger integral equation:
\begin{equation}
G_{\QEDL}(\vec r) = \int_V d^3 r' G_0(\vec r - \vec r\,')\left[\delta(\vec r\,') + 2\mu \VQEDL(\vec r\,')G_{\QEDL}(\vec r\,') \right]
\label{eq:LS}
\end{equation}
where $G_0(\vec r)$ is the Helmholtz solution with $e^2=0$ given in Eq.~\eqref{eq:helmholtzgreens}.  Equation~\eqref{eq:LS} can be expanded through first order in $e^2$ giving
\begin{eqnarray}
G_{\QEDL}(\vec r) &\approx& G_0(\vec r)
 + \int_V d^3 r' G_0(\vec r - \vec r\,')2\mu \VQEDL(\vec r\,')G_0(\vec r\,').
\label{eq:LS_1}
\end{eqnarray}

Now we must determine the singular and regular parts of the right-hand side of Eq.~\eqref{eq:LS_1}.  We will determine the singular parts first.  This can be done in position space where we argue that for small $r$ the finite-volume solution $G_0(\vec r,E)$ can be approximated by its infinite-volume counter part:
\begin{eqnarray}
G^\infty_0(\vec r) &=&  \int d^3 k \frac{e^{i\vec q \cdot \vec r}}{(2\pi)^3} \frac{1}{-\vec q\,^2+k^2} 
\label{eq:Yukawa0} \\
&=& - \frac{1}{4\pi}\frac{e^{ikr}}{r}\,.
\label{eq:Yukawa1}
\end{eqnarray}

The singular part of the zeroth-order term in Eq.~\eqref{eq:LS_1} is given immediately by Eq.~\eqref{eq:Yukawa1}.   However, the singular part of the first-order term can also be found using this same expression for $G^\infty_0(\vec r)$ by evaluating:
\begin{eqnarray}
\int_V d^3 r' G_0(\vec r - \vec r\,')\VQEDL(\vec r\,')G_0(\vec r\,') &&
\label{eq:singular0} \\
&& \hskip -2.0 in \approx \int_V d^3 r' G^\infty_0(\vec r - \vec r\,')V_C(\vec r\,')G^\infty_0(\vec r\,')
\label{eq:singular1} \\
&& \hskip -2.0 in = \frac{e^2}{(4\pi)^3} \int d^3 r' \frac{e^{ik|\vec r-\vec r\,'|}}{|\vec r-\vec r\,'|}
                              \frac{e^{ik r'}}{(\vec r\,')^2}
\label{eq:singular2}  \\
&& \hskip -2.0 in \approx \frac{e^2}{32\pi^2} \int_0^\infty d r' e^{ik r'} \int_{-1}^1 \frac{d\cos(\theta)}
                                                   {\sqrt{r^2+{r'}^2 -2rr'\cos(\theta)}} \\
\label{eq:singular3}
&& \hskip -2.0 in \approx \frac{e^2}{(4\pi)^2} \int_0^\infty d r' e^{ik r'}
\left[ \frac{1}{r}\theta(r-r') + \frac{1}{r'}\theta(r'-r)\right]
\label{eq:singular4} \\
&& \hskip -2.0 in \approx \frac{e^2}{(4\pi)^2}\left( \ln(1/kr) + \cdots\right)
\label{eq:singular5}
\end{eqnarray}
where the dots represents terms that are constants or of higher order as $r\to0$.  Since we are interested in the singularity at $r=0$, we need not be concerned about the large $r'$ region in the integral appearing above.  This behavior can be controlled, for example, by giving $k$ a small, positive imaginary part.

Finally we must determine the regular part, specifically the limit at $r=0$ of the right hand side of Eq.~\eqref{eq:LS_1} after the singular part determined above has been removed.  Now, the behavior of these functions in the entire periodic volume becomes important and the simple, infinite-volume formulae used in the above discussion of the logarithmic singularity are not adequate.  L\"uchser has done this for the $G_0(\vec r)$ term in  Eq.~\eqref{eq:LS_1}
and found:
\begin{eqnarray}
\lim_{r\to 0} G_0(\vec r) &\sim& -\frac{1}{4\pi r} - \left.\frac{1}{4\pi^2L}
                              \sum_{\vec n}\frac{1}{\left[(\vec n)^2 - \left(\frac{kL}{2\pi}\right)^2\right]^s}\right|_{s=1} 
\label{eq:reg0} \\
   &=& -\frac{1}{4\pi r} - \sqrt{4\pi}\frac{1}{4\pi^2L}Z_{00}\left(1,q^2\right),
\label{eq:reg1}
\end{eqnarray}
where $q= kL/2\pi$ and the evaluation at $s=1$ in Eq.~\eqref{eq:Z00} is described in  Ref.~\cite{Luscher:1990ux}.  We must now make a similar evaluation of the right most term in Eq.~\eqref{eq:LS_1}.  

In analogy with Eq.~\eqref{eq:reg1}, we define the regular part of the $e^2$ correction to the function $G_{\QEDL}(\vec r)$
as
\begin{eqnarray}
Y(q) &=& \lim_{r\to0} \sum_{\vec n}\sum_{\vec m\ne0} 
\frac{e^{i\vec n \cdot \frac{2\pi \vec r}{L}}}{(2\pi)^4\pi} \frac{1}{(\vec n)^2 - q^2}
\frac{1}{\vec m\,^2} \frac{1}{(\vec n-\vec m)^2 - q^2} - \frac{1}{4\pi}\ln(1/kr).
\label{eq:Y-def}
\end{eqnarray}
The relatively mild logarithmic singularity allows a direct numerical evaluation $Y(q)$.  

Next, following L\"uscher we introduce a Coulomb function $\phi_C(E)$ such that the combination:
\begin{equation}
\cos\left(\phi_C(E')\right)\mathcal{F}_0(k'r) - \sin\left(\phi_C(E')\right)\mathcal{G}_0(k'r)
\label{eq:phi_def}
\end{equation}
contains the same ratio of regular to singular parts as we have found in $G_{\QEDL}(E)$.  Equating the ratio of regular divided by singular parts in the limit of small $r$ in Eqs.~\eqref{eq:phi_def} and \eqref{eq:LS_1} we obtain:
\begin{eqnarray}
\frac{(\frac{2\pi\eta}{e^{2\pi\eta}-1})^{\frac{1}{2}}\,k'r\Bigl[ \cos(\phi_C) 
- \frac{e^{2\pi\eta}-1}{2\pi\eta}\sin(\phi_C) 2\eta'\left\{\mathrm{Re}\left[\psi(i\eta')\Bigr] + 2\gamma_E -1\right\}\right]}
{-\sin(\phi_C)\left[ (\frac{e^{2\pi\eta}-1}{2\pi\eta})^{\frac{1}{2}}
\Bigl(1+  2\eta' k'r \ln(k'r)\Bigr)\right]} &&
\label{eq:phi_first0} \\
&&\hskip -2.0 in 
= \frac{-\frac{1}{4\pi^2L}\sqrt{4\pi}Z_{00}(1,\frac{kL}{2\pi})+\frac{2\mu e^2}{4\pi} Y\left(\frac{kL}{2\pi}\right)}{-\frac{1}{4\pi r} + \frac{2\mu e^2}{(4\pi)^2} \ln(1/kr)}
\nonumber
\end{eqnarray}
where to determine the left-hand side we have used Eqs.~\eqref{eq:F-small_r} and \eqref{eq:G-small_r} while the 
right-hand side has been determined from Eqs.~\eqref{eq:reg1} and \eqref{eq:Y-def}.  In contrast with the zeroth order in $e^2$ case, the singular solution $\mathcal{G}_0$ at order $e^2$ contains a regular term which must be included.  To allow a direct comparison of Eq.~\eqref{eq:phi_first0} and Eqs.~\eqref{eq:F-small_r} and \eqref{eq:G-small_r} we have not expanded this equation to first order in $e^2$.

Recognizing that $\eta$ with $\mu e^2/k$ we can simplify in Eq.~\eqref{eq:phi_first0} by expanding to first order in $\eta$ to obtain a final formula for $\phi_C(E)$ to order $e^2$:
\begin{eqnarray}
\cot\bigl(\phi_C(E')\bigr) &=& -(1+\pi\eta)\frac{1}{\pi^{\frac{3}{2}}}\left(\frac{2\pi}{k'L}\right)
                          Z_{00}\left(1,\frac{kL}{2\pi}\right) 
\nonumber \\
&&\hskip 0.5 in      +8\pi\eta Y\left(\frac{kL}{2\pi}\right) 
                                      +2\eta\Bigl\{\mathrm{Re}\left[\psi(i\eta)\right] + 2\gamma_E -1\Bigr\}
\label{eq:phi_first1} 
\end{eqnarray}
For simplicity, we have not expanded the relation between the primed and unprimed variables to first order in $e^2$.

We can summarize the results of this appendix by stating the relation between the energy eigenvalue of a finite-volume energy eigenstate and the corresponding infinite-volume non-relativisitic $s$-wave scattering phase shift $\bar\delta_0(E')$ given by the quantization condition:
\begin{equation}
\cot(\bar\delta_0(E')) + \cot(\phi_C(E')) =0\,,
\label{eq:CQ}
\end{equation}
where $E'=E+\Delta E$ where $\Delta E$ is given in Eq.~\eqref{eq:DeltaE}.  Equation~\eqref{eq:CQ} is accurate up to but not including terms of order $(r_0/L)^3$ when the scattering potential $V_S$ vanishes outside the radius $r_0$.

\section{Numerical tests of \qedl finite-volume quantization}
\label{sec:QEDL-numerics}

After identifying the $O(\frac{r^2}{L^3})$ correction to the \qedl potential, one could become concerned about the size of such corrections. In order to understand the efficacy of the $\QEDL$-modified finite-volume quantization condition, it is prudent to work with a model system where the phase shift is known exactly. In such a system, one can attempt to discern the size of the effects of neglected $e^4$ terms and of the neglected inverse powers of $L$ by comparing with the precisely known result. This additional step is useful before embarking on a costly lattice QCD calculation comparing the truncated Coulomb and \qedl results in order to gauge expectations for accuracy. In this Appendix, we perform a numerical calculation of the phase shift in a model theory and determine the residual structure-dependent error introduced by the unphysical $O(e^2 L^{-3})$ term.

\subsection{$\delta$-shell potential}

Quantum mechanics with a $\delta$-shell potential,
\beq
V({\vec r}) = V_0\, \delta( |{\vec r}| - r_0 )\,,
\eeq
is a sufficiently simple system to solve both analytically in infinite volume and numerically in a finite volume. The $\delta$-shell potential has the feature necessary for application of L\"uscher's quantization condition, as well as the $\QEDL$-modified finite-volume quantization condition, that there exists an exterior region of $|{\vec r}| > r_0 $ beyond which the interaction does not exist. In the numerical finite-volume studies, the interaction range will be kept such that $r_0 < L/2$. 

In infinite volume, within the regions interior and exterior to the $\delta$-shell, the radial component of wave function will behave as a free particle and a solution to the spherical Bessel equation. In the exterior, the wave function will be a linear combination of the regular and irregular spherical Bessel functions
\beq\label{eq:exterior}
\psi(r > r_0) = k A j_l(kr) - k B n_l(kr)\,,
\eeq
where phase shift is found through the ratio $B/A = \tan\delta_l$. Since the wave function must remain finite, in the interior region the wave function will contain only the regular solution
\beq\label{eq:interior}
\psi(r < r_0) = k j_l(kr)\,,
\eeq
up to an arbitrary normalization.

For a $\delta$-shell potential, the wave function must be continuous at the shell and its derivative must have a discontinuity proportional to the value of the wavefunction on the shell. Applying these conditions, the phase shift is found to be
 \beq
 \tan \delta_l = \frac{\frac{V_0}k j_l^2(kr_0)}{(kr_0)^{-2} -\frac{V_0}k j_l(kr_0)n_l(kr_0)}\,,
 \eeq
 where the first term in the denominator comes from the Wronskian of these functions $n_l'(z)j_{l}(z) - n_{l}(z)j_l(z) = z^{-2}$.
 For the $s$-wave phase shift, this equation reduces to the textbook result:
 \beq\label{eq:inf_phase_no_coul}
 \tan \delta_0 = \frac{ \frac{V_0}k \sin^2(kr_0)}{1 + \frac{V_0}k \sin(kr_0)\cos(kr_0)}\,.
 \eeq

\subsection{Combined $\delta$-shell and Coulomb potential}

An identical derivation can be performed in the presence of the Coulomb potential, where the spherical Bessel functions are replaced by the regular, $F_l(\eta,z)$, and irregular, $G_l(\eta,z)$, Coulomb wavefunctions replacing $j_l(z)$ and $-n_l(z)$ respectively. These wavefunctions are related to the confluent hypergeometric function $U$ through their relation to the Coulomb analogue to the Hankel functions
\bea
H_l^{\pm}(\eta, kr) &=& G_l(\eta,kr) \pm i F_k(\eta,kr) \nonumber \\
&=& e^{\pm i \Theta(kr)} (\mp 2i kr)^{1+l \pm i \eta} U(1+l \pm i\eta, 2l +2, \mp 2ikr) \,,
\eea
where $\Theta(z) = z - l\pi/2 + \sigma_l(\eta) - \eta \ln(2z)$ and $\sigma_l(\eta) = {\rm arg} \,\Gamma(1+l+i\eta)$ is the Coulomb phase shift. .  \\

The exterior and interior wavefunction are parameterized in the just as in~\eqref{eq:exterior} and~\eqref{eq:interior} and are subject to the same boundary conditions at the radius of the $\delta$-shell. Using the Wronskian of the Coulomb wavefunctions $G_l(z)F_{l}'(z) - G_{l}'(z)F_l(z) = 1$, the phase shift in the presence of the Coulomb interaction will be given by
\beq\label{eq:inf_phase_with_coul}
 \tan \delta_l = \frac{\frac{V_0}k F_l^2(\eta, kr_0)}{1 +\frac{V_0}k F_l(\eta,kr_0)G_l(\eta,kr_0)}\,.
\eeq\\
By comparing numerical results with this analytical result, the efficacy of the $\QEDL$-modified finite-volume quantization condition can be tested. 

\subsection{Computational strategy}

In the remainder of this appendix we will calculate numerically the finite-volume energies for the combined $\delta$-shell and \qedl potentials, apply the \qedl finite-volume quantization condition given in Eq.~\eqref{eq:CQ} and then compare the results for the scattering phase shift with the exact infinite-volume values given by Eq.~\eqref{eq:inf_phase_with_coul}.  Our numerical approach is to determine the finite-volume energies and wave functions to order $e^0$ and then to use first-order perturbation to determine the $O(e^2)$ correction.

To solve the Schr\"odinger equation with $\delta$-shell potential alone in finite volume, again the wave function is parameterized in the regions interior and exterior to the $\delta$-shell. The wave function in the interior region is the same as given in Eq.~\eqref{eq:interior}, but the exterior region must satisfy the periodic boundary conditions. The exterior wavefunction is proportional to the finite-volume Green's function given in Eq.~\eqref{eq:helmholtzgreens}. The boundary conditions at the $\delta$-shell leads to the quantization condition
\beq\label{eq:fv_quant}
\left[ 1 - V_0 r_o - k r_0 \cot(kr
_0)\right] G_{0,0}(r_0;k) + r_0 \frac{\partial G_{0,0}}{\partial r}(r_0;k) = 0 \,.
\eeq
where $G_{0,0}(r)$ is the $s$-wave component of the zeroth-order Green's function $G_0(\vec r)$ defined in Eq.~\eqref{eq:helmholtzgreens}:
\begin{equation}
G_{0,0}(r) = \frac{1}{4\pi}\int d\Omega_{\hat r} G_0(r\hat r).
\label{eq:s-wave-projection}
\end{equation}
We will impose no conditions on higher partial waves with $l\ge 4$ making the usual assumption that the $\delta$-shell phase shifts for $l\ge 4$ can be neglected so that the regular behavior of $G_{0,l}(r)$ needs no modification.  Thus, this approximation is made in both our quantization condition and our numerical work and hence is not tested in their comparison.  The
zeros of Eq.~\eqref{eq:fv_quant} will be determined numerically.  The
numerical evaluation of $G(r;k)$ is described in Appendix
~\ref{app:ewald} and uses the Ewald summation technique to
dramatically improve the convergence of the sums in
Eq.~\eqref{eq:helmholtzgreens}. From those energy levels and the
corresponding wavefunction, the \qedl potential will be included using
first-order perturbation theory by calculating $\langle \psi | V_{{\rm
QED}_{\rm L}} | \psi \rangle$ numerically, as described in
Sec.~\ref{sec:numerical_delta}.

\subsection{Ewald summation}\label{app:ewald}
An efficient numerical evaluation of the Helmholtz Green's function in three dimensions is a non-trivial task. In its typical representation, Eq.~\eqref{eq:helmholtzgreens}, as a sum in momentum space
the series is slowly convergent. If one truncates the sum to only contain terms with ${\vec{p}} < p_{\rm max}$, then the leading neglected terms will be proportional to $p_{\rm max}^{-2}$, but there will be approximately $p_{\rm max}^{2}$ of them. The large number of these terms will create substantial contributions even if their individual summands are small and will also increase the computational costs as $p_{\rm max}$ is raised. A representation as a sum over the periodic images in position space can be found using Poisson's summation formula
\beq\label{eq:helm_pos}
G_0({\vec r};k) = \frac{1}{4\pi} \sum_{\vec n}  \frac{e^{i k | \vec r - \vec nL|}}{| \vec r - \vec nL|}
\eeq
which is even less convergent.\\

Variants of the Helmholtz Green's function are necessary in a wide range of applications in physics and engineering, such as determining the wave function of electrons within a quasi-periodic crystal. The evaluation of these functions has been expedited with a method called ``Ewald summation''~\cite{PhysRev.124.1786}. Ewald summation breaks the series into two components which represent the ``near'' contributions, which come from contributions of periodic images close to $\vec r$ and are evaluated in the position space representation, and ``far'' contributions, which come from the more distant periodic images and are evaluated with the momentum space representation~\cite{PhysRev.124.1786}. As will be seen, the rate of convergence will depend on the cutoff between these two regions. If the cutoff were taken to either extreme that all points were in one of these regions, the rate of convergence will blow up as anticipated. \\

To perform Ewald summation, the summand in Eq.~\eqref{eq:helm_pos} is rewritten as an integral 
\beq
 \frac{e^{ikr}}{4 \pi r} = \frac{1}{2 \pi^{3/2}} \int_\Gamma e^{-r^2 \xi^2} e^{k^2/\xi^2} d\xi \,,
\eeq
where $\Gamma$ is a particular contour from 0 to $\infty$ with some constraints on how it approaches those limits~\cite{PhysRev.124.1786}. First $\Gamma$ must leave the origin with a negative angle $\phi$ in the limits $0> \phi > -\frac \pi 4$, second it returns to the real axis at $\xi_0$, and then finally, continues on the real axis to $\infty$. This integral is then broken into two regions, from 0 to $\xi_0$ and from $\xi_0$ to $\infty$. The first of these regions will be evaluated in momentum space and the second in position space
\bea\label{eq:ewald}
G_0({\vec r};k) = G_1({\vec r};k) + G_2({\vec r};k) = 
\frac{1}{L^3} \sum_p \frac{e^{i{\vec p}\cdot {\vec r}}e^{\frac{k^2 - p^2}{4 \xi_0}}}{k^2 - p^2}  + \nonumber \\
\frac{-1}{4\pi}\sum_n {\rm Re}\left[ e^{i | {\vec r} + {\vec n}L | k}  {\rm erfc}\left(| {\vec r} + {\vec n}L | \xi_0 + \frac{i k}{2\xi_0}\right)\right] \,.
\eea
These sums converge significantly faster than the original sum. The first converges as $e^{-p^2 /4 \xi_0^2}$ and the second as $e^{-n^2L^2 \xi_0^2}$. The original momentum space sum is the special case of $\xi_0=0$ and the position space sum is reached as $\xi_0 \to \infty$ which both demonstrate the slow convergence. Using the optimal value of $\xi_0 = 3^{\frac 14}\sqrt{\frac{\pi}L}$~\cite{850366}, these sums can converge to sufficient accuracy with just a few terms. \\

These summation methods will also be necessary for numerically evaluating the \qedl potential efficiently. Unlike in a lattice application, with a UV regulator, the standard summation definition \qedl potential is also slowly convergent. For the numerical work, the \qedl potential will be evaluated using Eq.~\eqref{eq:qedlcoul}.

\subsection{Numerical analysis of $\delta$-shell potential}
\label{sec:numerical_delta}

The numerical analysis of the $\delta$-shell potential is performed for two case.   In one case the constant $V_0=10$, making the effects of the potential large compared to the kinetic energies studied.  For the second we choose $V_0=1$.  The radius of the $\delta$-shell is held fixed to 1. The size of the finite volume $L$ is varied from 4 to 8 to generate many energy levels.

\begin{figure}[!htp]
\centering
\includegraphics[width=0.48\textwidth]{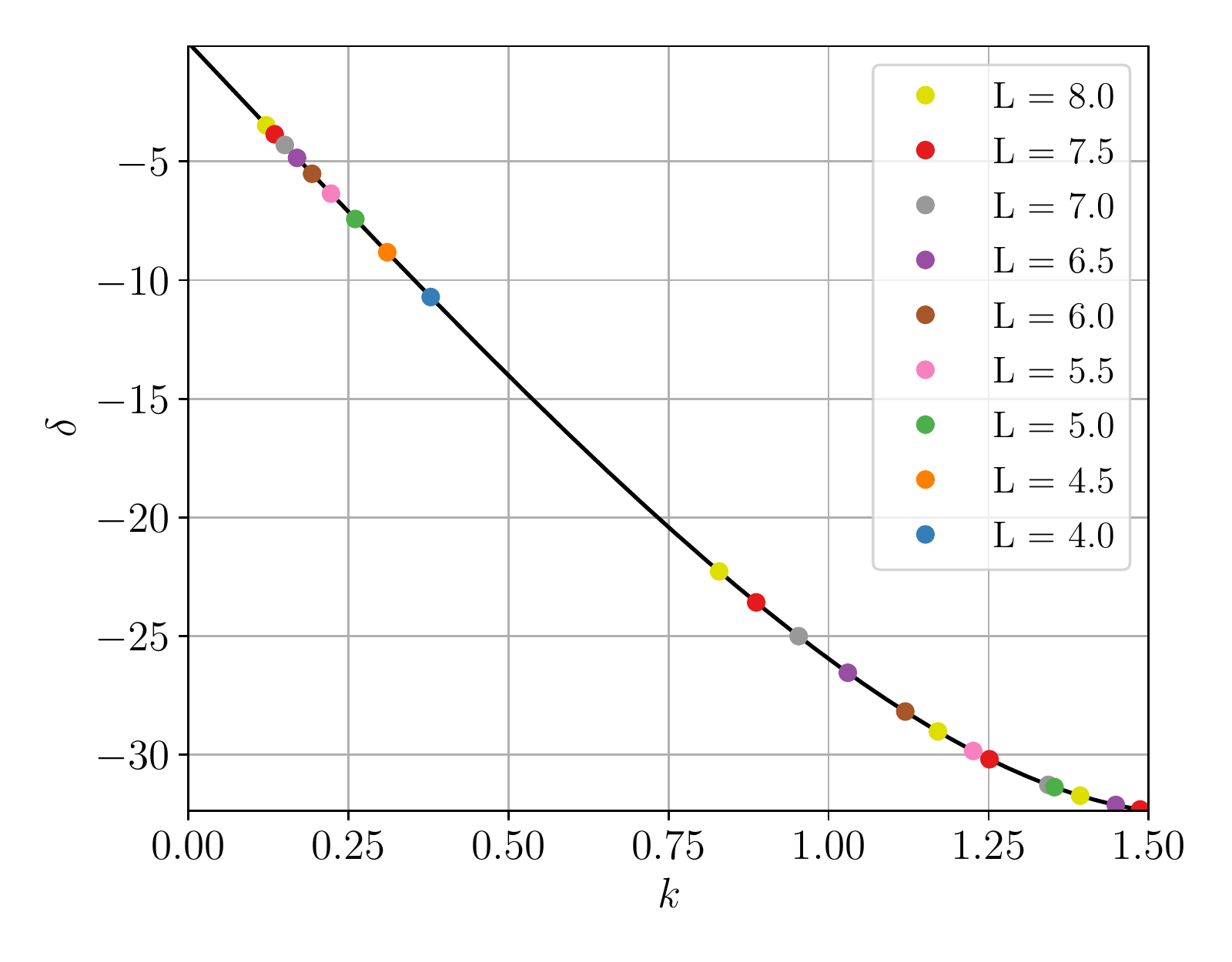}
\includegraphics[width=0.48\textwidth]{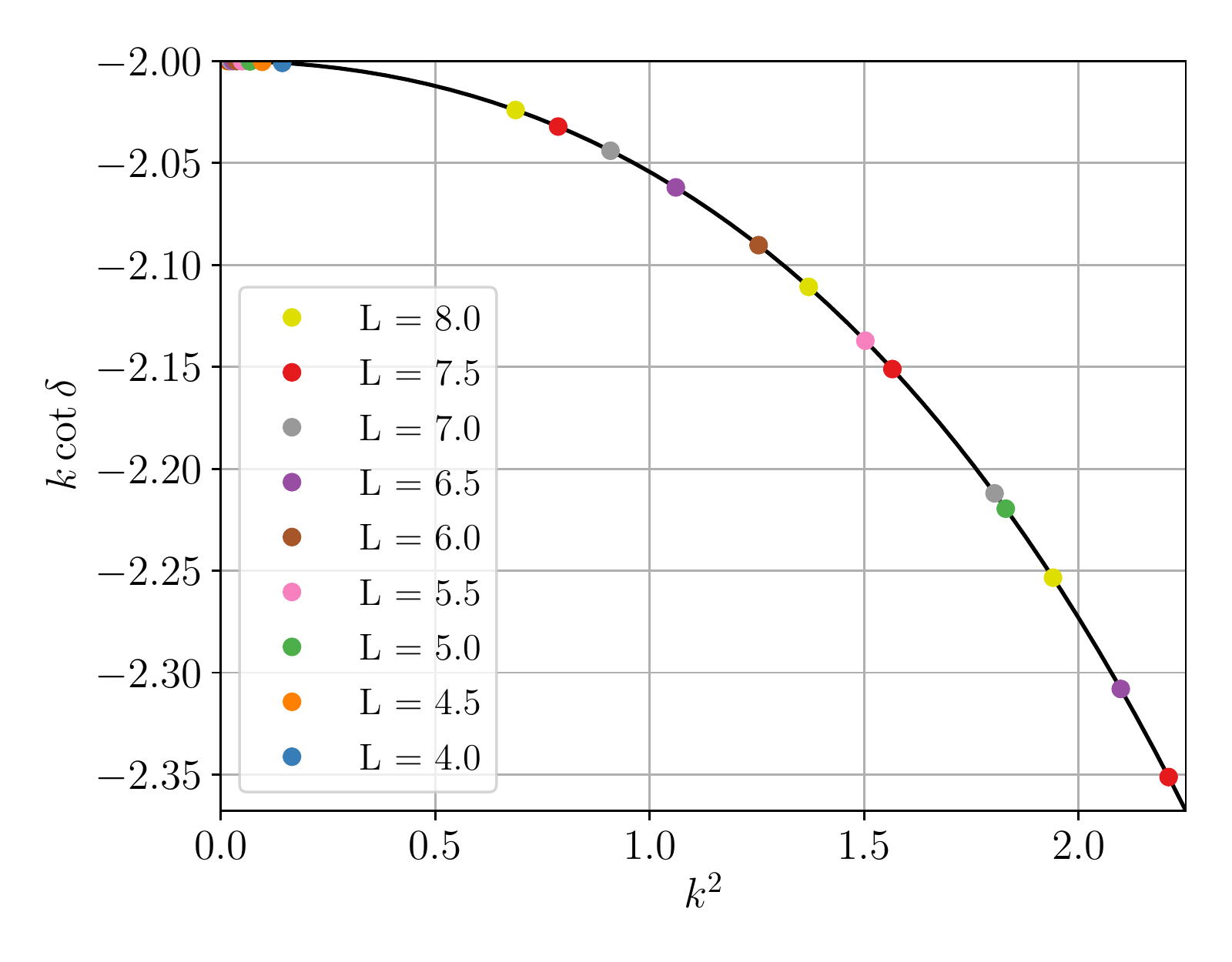}
\includegraphics[width=0.48\textwidth]{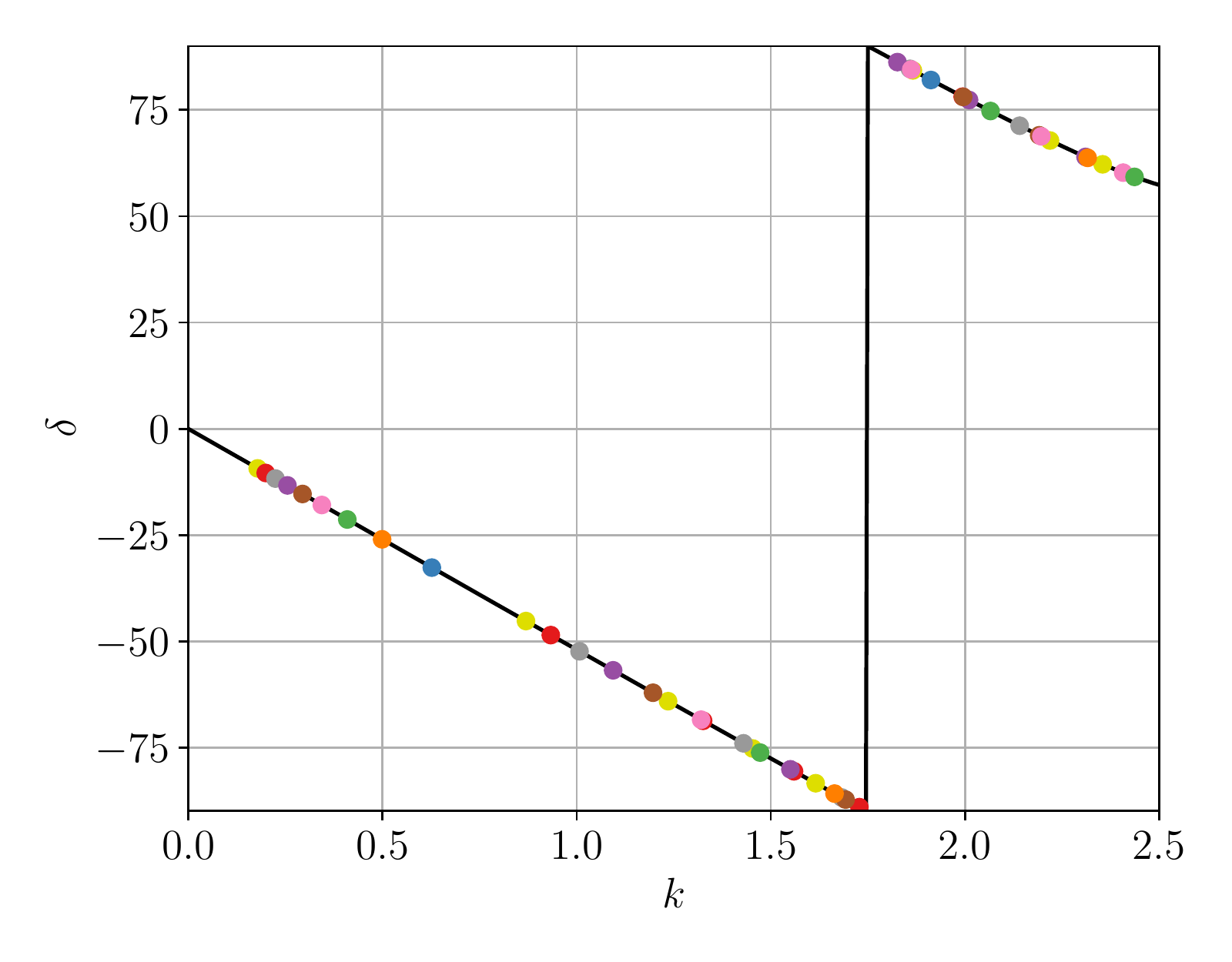}
\includegraphics[width=0.48\textwidth]{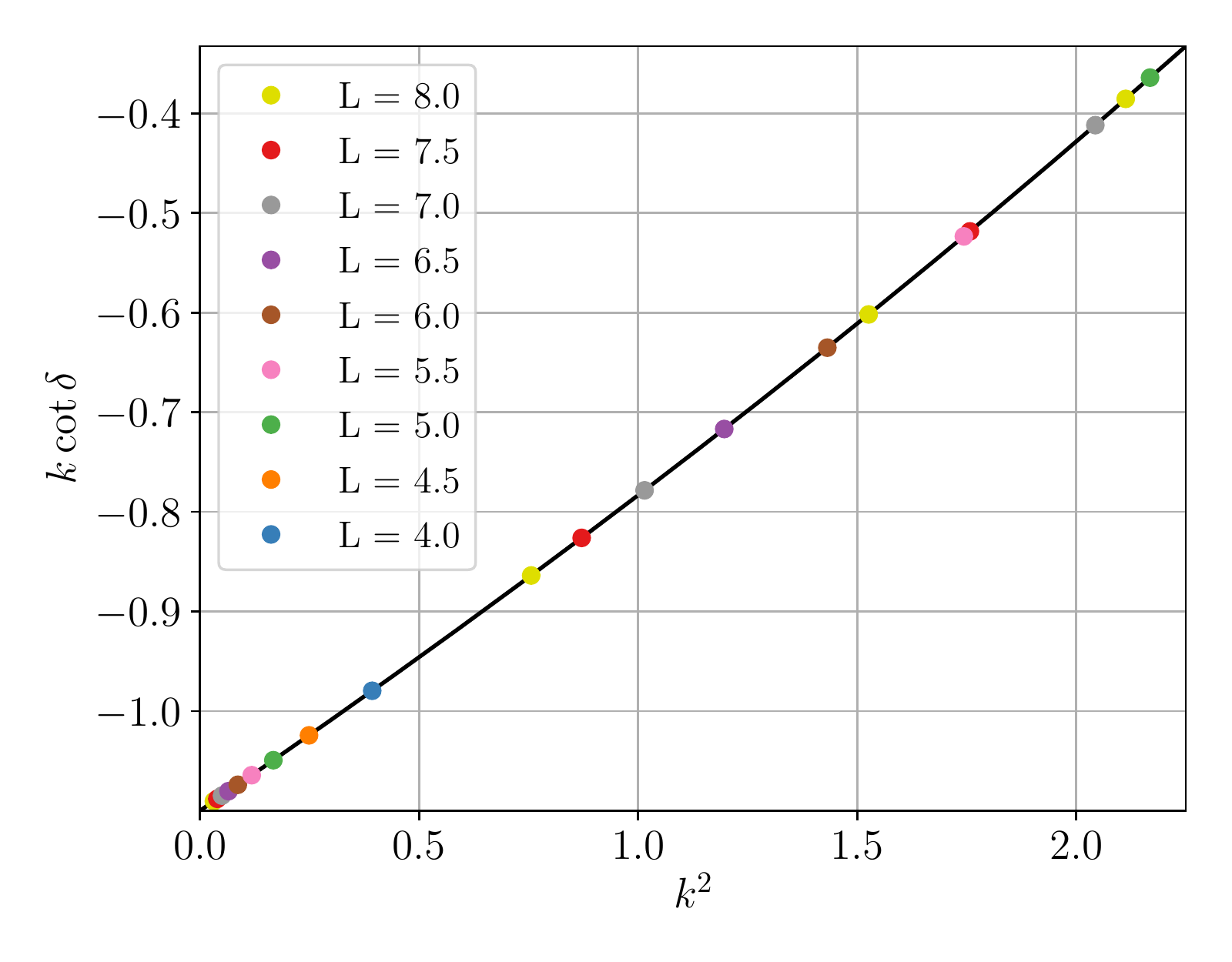}
\caption{\label{fig:delta_shell_no_coul} The scattering phase shift $\delta$ (Left) and $k\cot\delta$ (Right) for the $\delta$-shell potential. The strength of the $\delta$-shell potential is 1 for the upper plots and 10 for the lower plots. The black curve represents the exact infinite-volume solution of Eq.~\ref{eq:inf_phase_no_coul}.}
\end{figure}

The energy levels are calculated by finding the lowest zeros of Eq.~\ref{eq:fv_quant} over a range of volumes. The phase shift without Coulomb interactions, determined through L\"uscher quantization, is shown in Fig.~\ref{fig:delta_shell_no_coul}. In both cases, the finite-volume quantization condition reproduces the true phase shift, shown as the curve in that figure, to high accuracy over the range of $k$ studied. The largest discrepancies are $O(10^{-11})$. For the case of $V_0=10$, $k\cot\delta$, being linear in $k^2$ can be well described by the first two terms of the effective range expansion, but for the weaker potential more terms would be required.

\begin{figure}[!htp]
\centering
\includegraphics[width=0.48\textwidth]{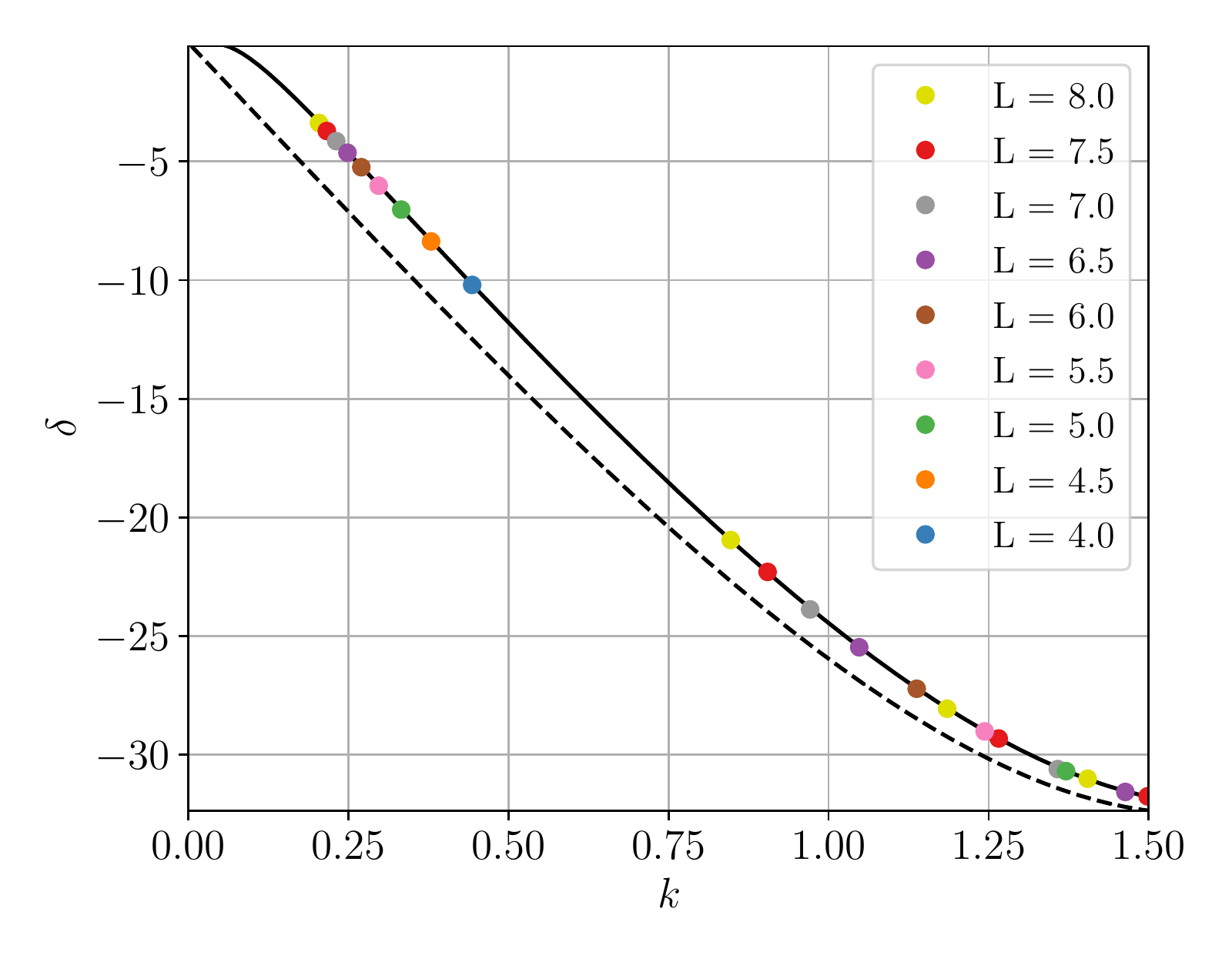}
\includegraphics[width=0.48\textwidth]{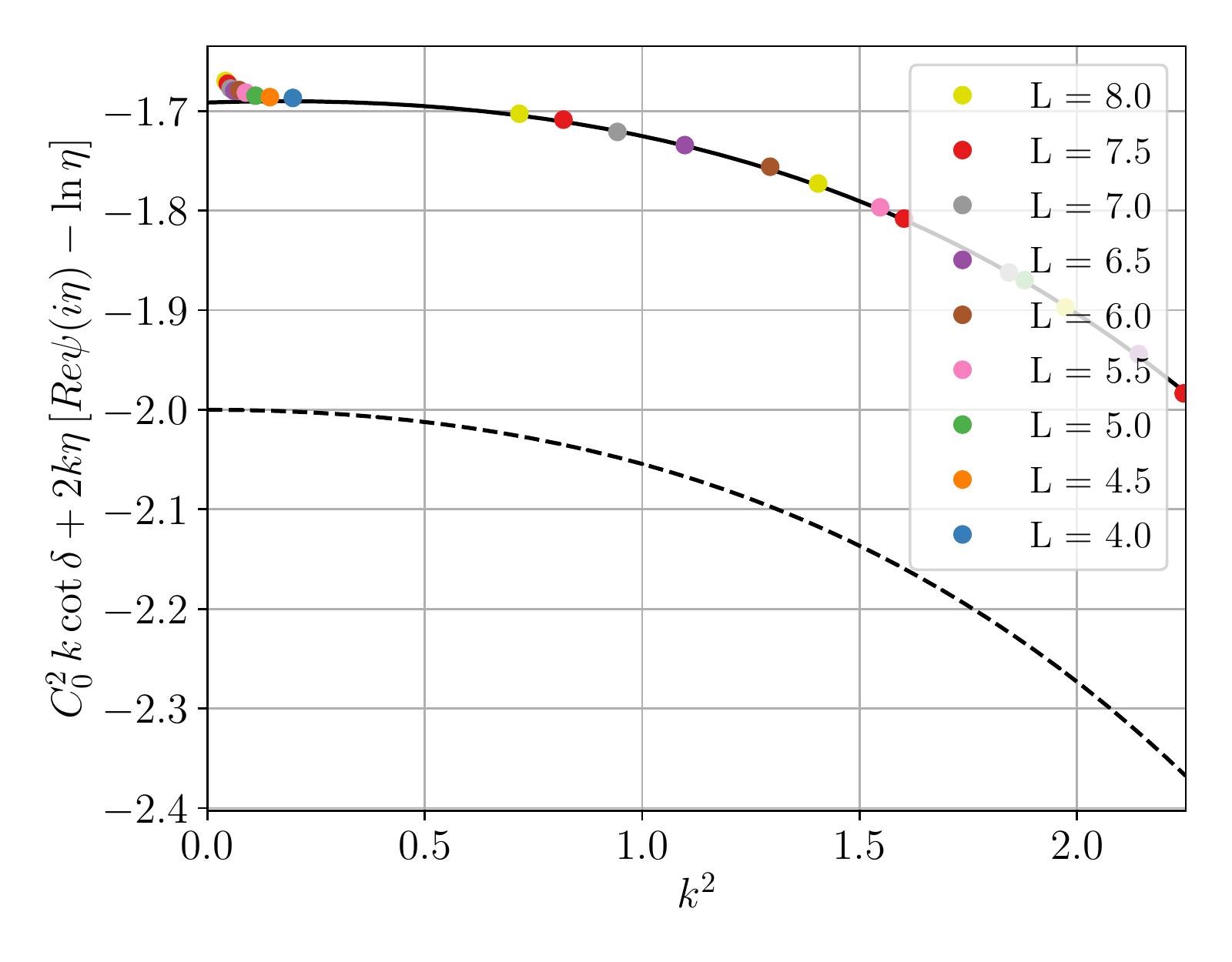}
\includegraphics[width=0.48\textwidth]{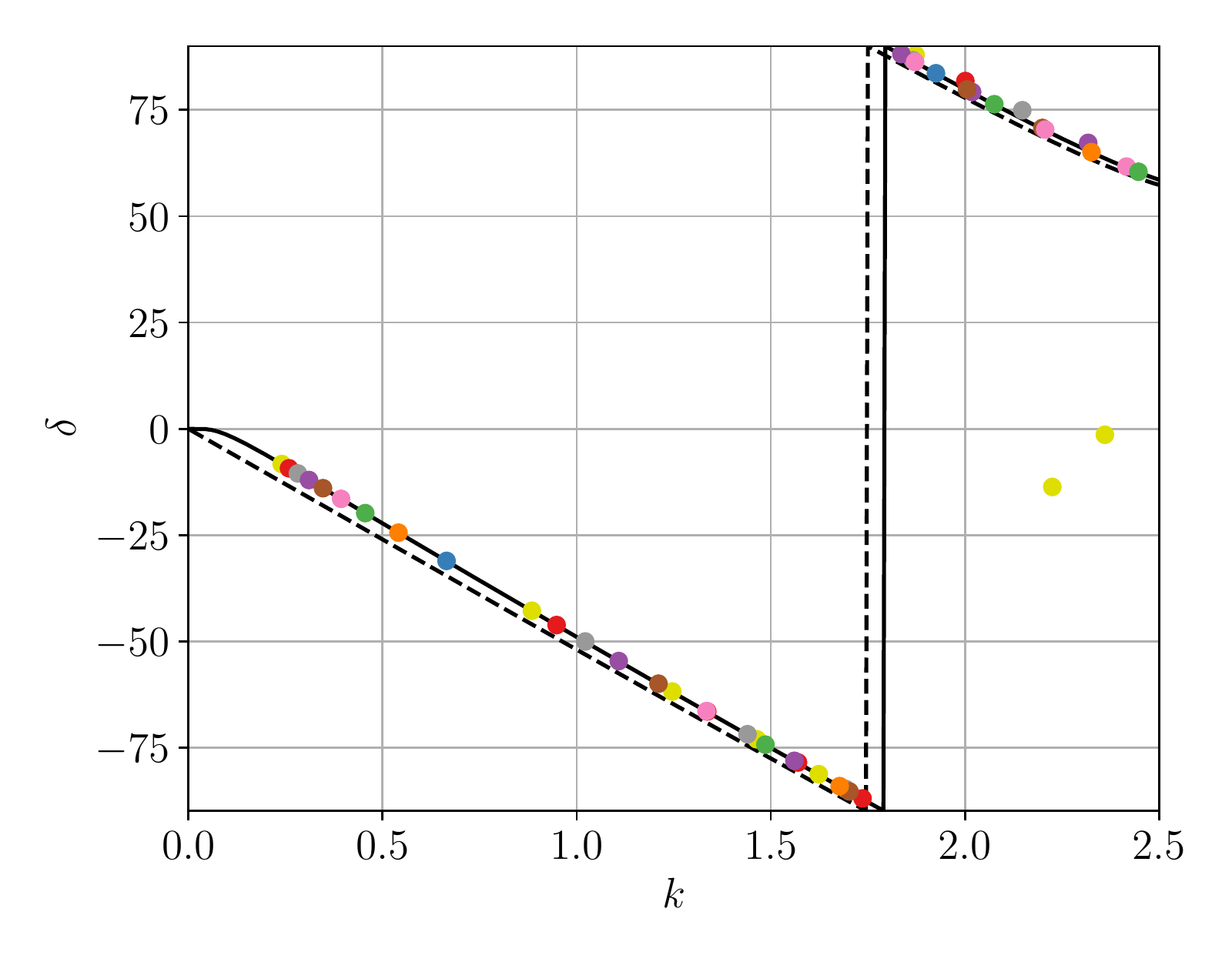}
\includegraphics[width=0.48\textwidth]{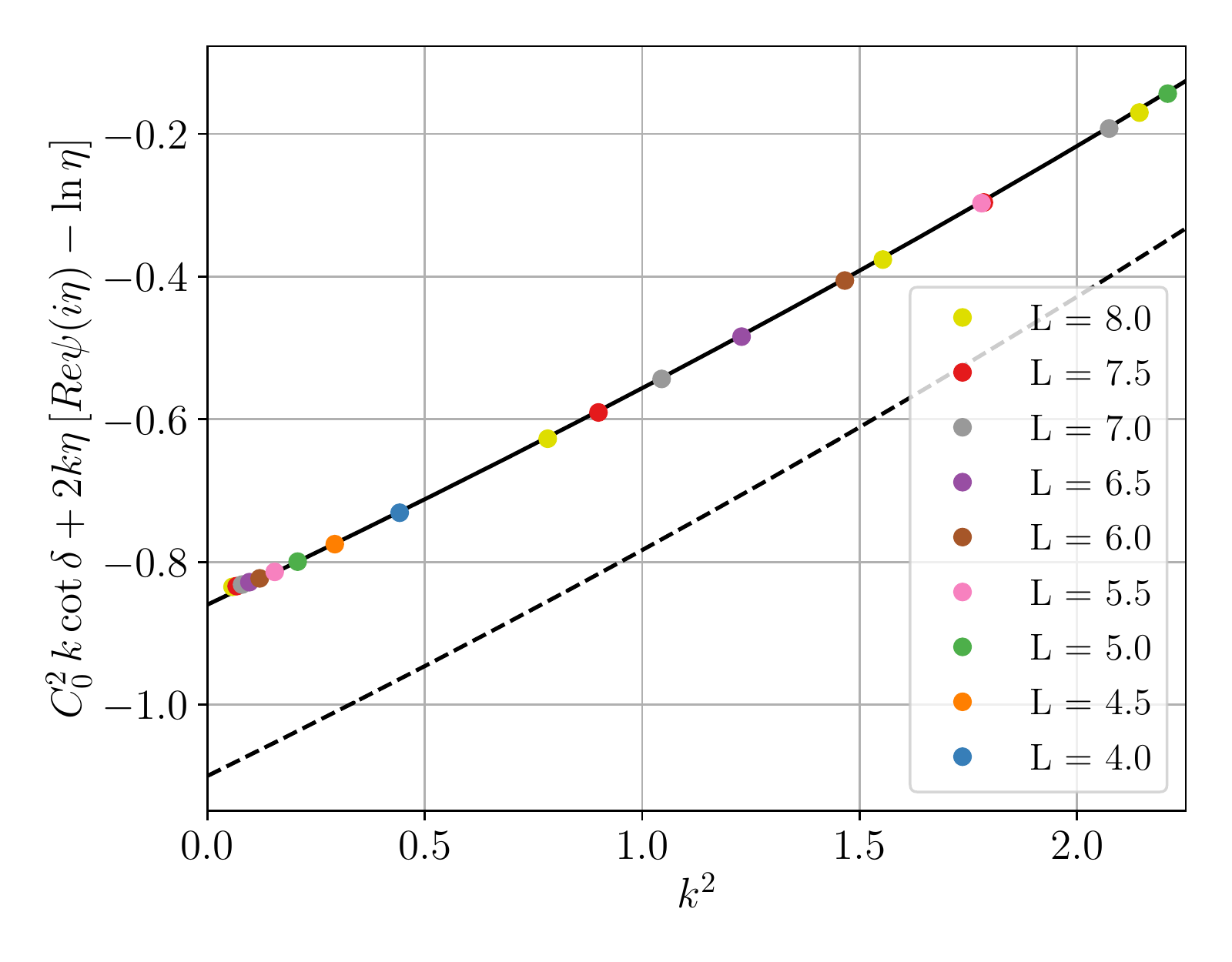}
\caption{\label{fig:delta_shell_with_coul} The scattering phase shift $\delta$ (Left) and $k\cot\delta$ (Right) for the combined $\delta$-shell and Coulomb potential. The strength of the $\delta$-shell potential is 1 for the upper plots and 10 for the lower plots. The continuous black curve represents the exact infinite-volume solution of Eq.~\ref{eq:inf_phase_with_coul} while the points are calculated using the finite-volume \qedl quantization condition developed here.  The dash black curve represents the exact infinite-volume solution without the Coulomb interaction given by Eq.~\ref{eq:inf_phase_no_coul}.}
\end{figure}

Perturbation theory is used to find the energy shifts caused by the inclusion of the \qedl potential. 
The unnormalized wavefunction corresponding to a finite-volume energy
eigenstate for the $\delta$-shell potential alone is approximately
\begin{equation}
    \psi({\vec r};k) = \begin{cases} \frac{\sin(kr)}{r} & r < r_0 \\
    \frac{\sin(kr_0)}{r_0 G_{0,0}(r_0;k)} G_0({\vec r};k) & r \ge r_0\,
    \end{cases}
\end{equation}
up to order $1/L^{2l}$ power-law corrections generated by ignoring contributions with
$l \ge 4$ in the interior region. Adding $V_{{\rm QED}_{\rm L}}$ as a
perturbation, the energy shift is calculated as $\Delta E =
\frac{\langle \psi | V_{{\rm QED}_{\rm L}} | \psi \rangle}{\langle
\psi | \psi \rangle}$ by analytically integrating the component of the
wavefunction for $r < r_0$ and then
performing a Monte-Carlo integration for the exterior region.
The value of the coupling is set to $e^2 = \frac{1}{2\mu}$, such that the effective Bohr radius,  $4\pi/(e^2\mu)$  is significantly larger than the size of the box. In the region of small $k$, the Sommerfeld parameter $\eta$ will begin to grow large and the perturbation theory used in the quantization condition begins to break down. Fig.~\ref{fig:delta_shell_with_coul} shows the results of the phase shifts from $O(e^2)$ energy levels using the finite-volume quantization condition of Eq.~\eqref{eq:CQ}. Even though the phase shift changes quite dramatically from the neutral case, the vast majority of the  phase shifts determined from the finite-volume quantization condition reproduce the infinite-volume result with sub-percent level accuracy. The lowest $k^2$ for the weak interaction case of $V_0=1$ differ by $O(1 \%)$ from the true infinite-volume non-perturbative result due to this breakdown of perturbation theory at large $\eta$. 

This study of a simple quantum mechanics problem implies that the structure-dependent effects of the unphysical $L^{-3}$ introduced by the \qedl potential on the phase shift are small for this case. Their contribution to the perturbative energy shift from the region within the radius of the strong interaction is, at their largest, $O(1.2\%)$ for $L=4$ of the contribution from the true Coulomb interaction in the same region. Na\"ively, one could expect that to be the size of the error in the phase shift. The relative error in the resulting phase shift for this case is $O(0.1\%)$. In fact, it appears the larger deviations from the infinite-volume non-perturbative result are from the truncation of perturbation theory in the finite-volume quantization condition of Eq.~\eqref{eq:CQ}, which grow largest at small $k$ where $\eta$ goes above 1. Of course, there is no guarantee the the structure-dependent effects of the $1/L^3$ term will be the same in an actual  QCD calculation, but analyzing this would require a full numerical lattice study comparing this \qedl approach to a more accurate method such as the truncated Coulomb potential approach presented in this paper.

\bibliography{references}
\end{document}